\documentclass[aps,prb,10pt,english]{revtex4-1}
\usepackage[T1]{fontenc}
\usepackage[utf8]{inputenc}
\usepackage[letterpaper]{geometry}
\usepackage{amsmath}
\usepackage{amssymb}
\usepackage{graphicx}
\usepackage{setspace}
\usepackage{esint}
\PassOptionsToPackage{normalem}{ulem}
\usepackage{ulem}
\usepackage[unicode=true,
bookmarks=false,
breaklinks=false,pdfborder={0 0 1},colorlinks=true]
{hyperref}
\hypersetup{
	final,linkcolor=blue,citecolor=blue,filecolor=magenta,urlcolor=blue}

\newcommand{\Tr}[0]{\textnormal{Tr}}

\emergencystretch=\maxdimen
\begin{document}
\title{Efficient choice of coloured noise in the stochastic dynamics of open
quantum systems}
\author{D. Matos$^{1}$, M. A. Lane$^{1}$, I. J. Ford$^{2}$, and L. Kantorovich$^{1}$}
\affiliation{\singlespacing{}$^{1}$Department of Physics, King's College London,
	Strand London, WC2R 2LS, United Kingdom}
\affiliation{\singlespacing{}$^{2}$Department of Physics and Astronomy, University
	College London, Gower Street, London, WC1E 6BT, United Kingdom}

\begin{abstract}
The Stochastic Liouville-von Neumann (SLN) equation describes the
dynamics of an open quantum system reduced density matrix coupled
to a non-Markovian harmonic environment. The interaction with the
environment is represented by complex coloured noises which drive
the system, and whose correlation functions are set by the properties
of the environment. We present a number of schemes capable of generating
coloured noises of this kind that are built on a noise amplitude reduction
procedure {[}Imai \textit{et al}, Chem. Phys. 446, 134 (2015){]},
including two analytically optimised schemes. In doing so, we pay
close attention to the properties of the correlation functions in
Fourier space, which we derive in full. For some schemes the method
of Wiener filtering for deconvolutions leads to the realisation that
weakening causality in one of the noise correlation functions improves
numerical convergence considerably, allowing us to introduce a well
controlled method for doing so. We compare the ability of these schemes,
along with an alternative optimised scheme {[}Schmitz and
Stockburger, Eur. Phys. J.: Spec. Top. \textbf{227}, 1929 (2019){]}, to reduce the growth
in the mean and variance of the trace of the reduced density matrix,
and their ability to extend the region in which the dynamics is stable
and well converged for a range of temperatures. By numerically optimising
an additional noise scaling freedom, we identify the scheme which
performs best for the parameters used, improving convergence by orders
of magnitude and increasing the time accessible by simulation.
\end{abstract}

\maketitle

\section{Introduction}

In open quantum systems, interactions between the system of interest
and its environment drive behaviours such as dissipation and decoherence
which are not found in isolation. These play a strong role in quantum
computing \citep{shor1995scheme} where the ability of the open system
to stay in a superposition of states is desirable, and also in quantum
thermodynamics \citep{weiss2012quantum}. Unfortunately, the very large
number of environmental degrees of freedom makes the treatment of
both the system and environment analytically and numerically challenging,
especially when there is strong coupling between them. For this reason,
existing methods tend to begin by taking the partial trace of the
full density matrix over the environment variables to obtain the reduced
density matrix of the system of interest. In particular, this is done
in the well-known Feynman-Vernon influence functional formalism where
the response of a bath coupled linearly to the open system is expressed
as a path integral over an infinite number of displaced harmonic oscillators
\citep{feynman2000theory}. Techniques which build on this method include
hierarchical equations of motion (HEOM) \citep{shao2004decoupling,yan2004hierarchical,yan2016stochastic}, hybrid stochastic HEOM \citep{zhou2005stochastic,zhou2008solving}, hierarchy of pure states \citep{suess2014hierarchy},
stochastic Schr\"{o}dinger equations \citep{orth2013nonperturbative},
quasiadiabatic path integrals \citep{makri1995tensor}, Stochastic
Liouville-von Neumann equations (SLNs) \citep{diosi1998non,stockburger2001non,stockburger2002exact,stockburger2004simulating},
and the Extended SLN (ESLN) equations method, which accounts for initial
thermalisation by the inclusion of an additional stochastic differential
equation (SDE) in imaginary time with imaginary time noises \citep{mccaul2017partition,mccaul2018driving,lane2020exactly}.
Importantly, none of these methods make the Markov assumption, where
the environment correlation times are taken to be negligibly short
compared to the characteristic timescales of the open system. The
Markov assumption has the physical interpretation that any information
dissipated from the system to the environment will never be returned,
i.e. the system-environment coupling is memoryless. Instead, the environment
is allowed to be fully non-Markovian, introducing a source of memory
to the system.

The SLN and ESLN methods, amongst others, are based on solving SDEs
with complex correlated (coloured) Gaussian noises. Beginning with
the seminal work of Grabert, Schramm and Ingold \citep{grabert1988quantum},
these methods evolve stochastic reduced density matrices via SDEs,
driven by the aforementioned noises, with the physical density matrix
being recovered by stochastic averaging over all realisations of these
noises. The advantage of these methods is that they are exact, non-perturbative,
and are in principle applicable to any temperature, system-environment
coupling strength, and any form of the spectral density.  In addition, with the recent development of the
ESLN, the system and its environment can be thermalised via the application
of an initial evolution in imaginary time, rather than being initialised
in a partitioned state \citep{diosi1998non,stockburger2001non,stockburger2002exact,stockburger2004simulating}.
The current work focuses on the properties of the noises and their
generation rather than on thermalisation or the properties of specific
physical dynamics, so we shall limit ourselves to the SLN rather than
the ESLN for simplicity. It is important to note that these methods
do not constitute an \textit{ad hoc} representation of the system
behaviour where the noises might have been introduced artificially
to model the environment. Instead, they have been derived rigorously
from an appropriate consideration of the whole system, consisting of both the open system and its environment, by means of elimination
of the environment using the path-integral method and a Hubbard-Stratonovich
transformation.

To simulate these SDEs, particular care must be taken when generating
the complex coloured noises, as the choice of a generation scheme
can significantly alter the statistical properties of the noises and
thus the system dynamics. A poor choice is characterised by a catastrophic
loss of trace preservation for the reduced density matrix, which requires
an exponentially large sample for convergence of the average. Making
this choice is not trivial since the correlation functions must be
satisfied with sensible decisions being made wherever there is freedom
to do so, especially for stronger coupling when the magnitude of the
noises is already large. In fact, one of the main conclusions of this
work is that generating noises which satisfy the desired correlation
functions is not enough to guarantee convergence or that the results
be physical, despite the correlation functions being the only formal
requirements of the theory on the noises.

In our previous work \citep{lane2020exactly}, one particular noise
generation scheme was used which produced well converged results as
a verification of the ESLN method, but here we generalise our procedure
and explore a number of possible noise generation schemes which all
create the desired correlation functions but produce results of different
convergence for the open system dynamics. We also optimise the scheme
to minimise the (erroneous) exponential growth of the trace, something
which has only recently been studied in any detail \citep{imai2015fmo,stockburger2019variance,lane2020exactly},
though with some inconsistencies in \citep{imai2015fmo} which we correct,
and compare our optimised scheme with a recently proposed alternative,
obtained independently via a different method and optimised subject
to different constraints \citep{stockburger2019variance}. By examining
the properties of the Fourier transforms of the desired correlations,
the properties of the different noises, and their effect on the system
dynamics, we arrive at a number of conclusions about noise generation
for SLN methods, where and why issues arise, and how to maximise the
possible duration (run time) of simulations before the stochastic
nature of the dynamics inevitably leads to numerical blow up and statistical
uncertainty.

For this purpose, we will use the spin-boson model as it is a relatively
simple model consisting of a two-level spin system surrounded by bosonic
degrees of freedom that describe the environment. This can naturally
be applied to qubits coupled to an environment \citep{duan1998reducing,costi2003entanglement,van2003engineering,kopp2007universal,cui2009non},
electronic energy transfer in biological systems \citep{imai2015fmo},
Josephson junctions \citep{makhlin2001quantum,liu2002theory,valenti2014switching},
cold atoms \citep{orth2010dynamics,orth2008dissipative} and solid-state
artificial atoms \citep{berns2008amplitude}. The spin-boson model
has already been considered previously by us in the context of the
ESLN \citep{lane2020exactly}.

Comparison with other methods mentioned in the Introduction is not within the scope of this paper, as it will be focusing only on the details of the noise generation within the SLN equation formalism. So, the purpose of the present paper is fourfold: (1) develop a general
scheme for noise generation for the SLN equation and propose a number
of possible choices for the scheme, including a fully optimised choice,
(2) demonstrate that these choices significantly alter the properties
of the noises with appropriate use of deconvolution methods \citep{smith1997scientist,hansen2002deconvolution}
where necessary, (3) examine in detail how different choices affect
the convergence properties of the results and the accessible run time
of simulation before blow up, comparing with other optimisation schemes
where possible \citep{stockburger2019variance,imai2015fmo}, and (4)
explain in detail why particular choices fail, referring to the properties
of the correlation functions themselves where necessary. More concretely,
in Sec. \ref{sec:theory} we briefly review the SLN formalism and
the spin-boson model, before introducing in Sec. \ref{sec:noise_generation}
our specific framework for noise generation and the possible choices
we have identified. Finally in Sec. \ref{sec:Results} we present
the results of the various noise generation schemes.

\section{Theory\label{sec:theory}}

\subsection{Stochastic Liouville-von Neumann equations}

Following the influence functional formalism of Feynman and Vernon
\citep{feynman2000theory}, we consider the standard setup of an open
quantum system with coordinates $q$ and Hamiltonian $H_{q}$ (that
may describe either an electronic or bosonic subsystem, or both) coupled
to an environmental heat bath of harmonic oscillators $i$ with masses
$m_{i}$, governed by a potential energy that is quadratic in the
oscillator displacement coordinates $\xi_{i}$. The coupling between
the open system and its environment is linear in the environment coordinates
but fully general in $q$, taking the form $-\xi_{i}f_{i}(q)$ for
the given coordinate $\xi_{i}$, with the $f_{i}(q)$ being arbitrary
functions of $q$. The full system Hamiltonian is thus
\begin{equation}
H_{\textnormal{tot}}(q,\{\xi_{i}\},t)=H_{q}(q,t)+\sum_{i}\frac{p_{i}^{2}}{2m_{i}}+\frac{1}{2}\sum_{ij}\Lambda_{ij}\xi_{i}\xi_{j}-\sum_{i}\xi_{i}f_{i}(q),\label{eq:open_system_H}
\end{equation}
where $p_{i}$ are momentum coordinates canonical to $\xi_{i}$, and
$\Lambda_{ij}$ is the force constant matrix of the bath. This is
a more general form of the Caldeira-Leggett Hamiltonian \citep{caldeira1983path}
since the environment coupling is a general function of $q$ rather
than being strictly bilinear.

In the SLN method the open system and environment density matrices
are initialised in a partitioned state where the full density matrix
$\boldsymbol{\rho}_{0}=\boldsymbol{\rho}_{\textnormal{tot}}(t_{0})$
is the tensor product of the open system density matrix $\boldsymbol{\rho}_{q}(t_{0})$
and that of its environment $\boldsymbol{\rho}_{\xi}(t_{0})$ at some
initial time $t_{0}$,
\begin{equation}
\boldsymbol{\rho}_{0}=\boldsymbol{\rho}_{q}(t_{0})\otimes\boldsymbol{\rho}_{\xi}(t_{0}).\label{eq:full_initial}
\end{equation}
In principle, the open system and its environment can be initialised
in the canonical equilibrium state using the ESLN formalism \citep{mccaul2017partition,lane2020exactly},
with the system and environment in thermal contact such that they
are fully thermalised. However, here we shall limit ourselves to the
partitioned initial state [Eq. (\ref{eq:full_initial})] and the SLN
method.

Tracing over the environment variables \citep{grabert1988quantum,stockburger2004simulating},
it is possible to obtain the Stochastic Liouville-von Neumann (SLN)
equation, an SDE which describes the evolution of a stochastic reduced
density matrix for the system driven by complex coloured noises, where
the physical reduced density matrix is obtained by taking the average
over a sample of many realisations of the dynamics. This SLN takes
the form
\begin{equation}
i\hbar\frac{d\boldsymbol{\rho}(t)}{dt}=\left[H_{q}(t),\boldsymbol{\rho}(t)\right]-\eta(t)\left[f(q),\boldsymbol{\rho}(t)\right]-\frac{\hbar}{2}\nu(t)\left\{ f(q),\boldsymbol{\rho}(t)\right\} ,\label{eq:dynamics}
\end{equation}
where $\boldsymbol{\rho}(t)$ represents the stochastic reduced density
matrix and the square (curly) brackets represent standard (anti-)commutators,
with the physical reduced density matrix given by $\boldsymbol{\rho}^{{ph}}(t)=\langle\boldsymbol{\rho}(t)\rangle$.
Here, $\eta(t)$ and $\nu(t)$ are the aforementioned complex coloured
noises, angle brackets $\left\langle \ldots\right\rangle $ represent
an average over the noises, $H_{q}(t)$ is the open system Hamiltonian
mentioned previously (which may depend explicitly on time), and $f(q)$
is the (universal) function which couples the system to the environmental
oscillators, assumed to be time independent.

The noises all have zero mean and are otherwise defined by their correlation
functions
\begin{equation}
\langle\eta(t)\eta(t^{\prime})\rangle=\hbar\int_{0}^{\infty}\frac{d\omega}{\pi}J(\omega)\coth\left(\frac{1}{2}\beta\hbar\omega\right)\cos\left(\omega\left(t-t^{\prime}\right)\right)\equiv K_{\eta\eta}(t-t^{\prime}),\label{eq:kernel_etaeta}
\end{equation}
\begin{equation}
\langle\eta(t)\nu(t^{\prime})\rangle=-2i\Theta(t-t')\int_{0}^{\infty}\frac{d\omega}{\pi}J(\omega)\sin\left(\omega\left(t-t^{\prime}\right)\right)\equiv K_{\eta\nu}(t-t^{\prime})=iR\left(t-t^{\prime}\right),\label{eq:kernel_etanu}
\end{equation}
\begin{equation}
\langle\nu(t)\nu(t^{\prime})\rangle=0,\quad\forall t,t^{\prime},\label{eq:zero_correlations}
\end{equation}
where $J(\omega)$ is the spectral density of the environment and
$\beta=1/k_{B}T$ where $T$ is the temperature of the environment.
From now on we set $\hbar=1.$ In this study we take $J(\omega)$
to be of the Drude form
\begin{equation}
J(\omega)=\omega\left[1+\left(\frac{\omega}{\omega_{c}}\right)^{2}\right]^{-2},
\end{equation}
where the cut off frequency $\omega_{c}$ controls the decaying character
of $J(\omega)$ at large $\omega$, and there is a hard cutoff such
that $J\left(\omega>\omega_{c}\right)=0$. To be explicit, the Drude form of the spectral density is used prior to a hard cutoff $\omega_{c}$ that specifies the maximum phonon frequency of the bath above which there is no contribution associated with higher frequencies.

\subsection{Spin-Boson Model}

Thus far, the system Hamiltonian $H_{q}$ has been kept fully general,
as has the form of the system-environment coupling, $f(q)$. We will
adopt the spin-boson Hamiltonian for our system of interest which,
in a basis of a generic two-state system, is
\begin{equation}
H_{q}(t)=\frac{1}{2}\Delta(t)\sigma_{x}+\frac{1}{2}\epsilon(t)\sigma_{z}=\frac{1}{2}\Delta(t)\left(\lvert0\rangle\langle1\rvert+\lvert1\rangle\langle0\rvert\right)+\frac{1}{2}\epsilon(t)\left(\lvert0\rangle\langle0\rvert-\lvert1\rangle\langle1\rvert\right).\label{eq:spin_boson_H}
\end{equation}
Here, $\sigma_{x,y,z}$ are the standard Pauli spin matrices with
$\sigma_{x}$ flipping the spin from one state to the other with tunnelling
strength $\Delta(t)$ and $\sigma_{z}$ biasing the energy of states
with magnitude $\epsilon(t)$. The system-bath coupling (previously
$f\left(q\right)$ in Eq. (\ref{eq:dynamics})) is $\alpha\sigma_{z}$,
where $\alpha$ is the coupling strength between the open system and
the environmental oscillators. Equation (\ref{eq:dynamics}) then becomes
\begin{equation}
i\frac{d\boldsymbol{\rho}(t)}{dt}=\left[H(t),\boldsymbol{\rho}(t)\right]-\alpha\eta(t)\left[\sigma_{z},\boldsymbol{\rho}(t)\right]-\frac{1}{2}\alpha\nu(t)\left\{ \sigma_{z},\boldsymbol{\rho}(t)\right\} .\label{eq:ESLN-method-1}
\end{equation}
 Finally, for the spin-boson Hamiltonian it is straightforward to
derive coupled SDEs for the $x$-, $y$- and $z$-spins and $\Tr{\boldsymbol{\rho}(t)}$
directly,
\begin{equation}
\frac{d\sigma_{x}(t)}{dt}=-\left[\epsilon(t)-2\alpha\eta(t)\right]\sigma_{y}(t)\label{eq:xspin}
\end{equation}
\begin{equation}
\frac{d\sigma_{y}(t)}{dt}=-\Delta(t)\sigma_{z}(t)+\left[\epsilon(t)-2\alpha\eta(t)\right]\sigma_{x}(t)\label{eq:yspin}
\end{equation}
\begin{equation}
\frac{d\sigma_{z}(t)}{dt}=\Delta(t)\sigma_{y}(t)+i\alpha\nu(t)\Tr{\boldsymbol{\rho}(t)}\label{eq:zspin}
\end{equation}
\begin{equation}
\frac{d\Tr{\boldsymbol{\rho}}(t)}{dt}=i\alpha\nu(t)\sigma_{z}(t).\label{eq:trace_evolution1}
\end{equation}
To be clear, these are expectation values of spins $\sigma_{x,y,z}(t)=\Tr\left(\sigma_{x,y,z}\boldsymbol{\rho}(t)\right)$
obtained from a single realisation of the stochastic reduced density
matrix. The physical expectation values would then be obtained by
the average over many such realisations, $\langle\sigma_{x,y,z}\left(t\right)\rangle$.

\section{Noise Generation Schemes\label{sec:noise_generation}}

The correlation functions given by Eqs. (\ref{eq:kernel_etaeta})-(\ref{eq:zero_correlations})
act as constraints on the noise generated, but do not uniquely define
them, leaving some freedom to specify the generation procedure.

For the purpose of considering different representations of the noises,
we adopt the most general form of the linear filtering ansatz \citep{oppenheim1999discrete},

\begin{equation}
\eta\left(t\right)=\int_{-\infty}^{\infty}dt^{\prime}\sum_{j}F_{j}f_{j}\left(t-t^{\prime}\right)x_{j}\left(t^{\prime}\right)\label{eq:general_eta}
\end{equation}

\begin{equation}
\nu\left(t\right)=\int_{-\infty}^{\infty}dt^{\prime}\sum_{j}G_{j}g_{j}\left(t-t^{\prime}\right)x_{j}\left(t^{\prime}\right),\label{eq:general_nu}
\end{equation}
where the $\left\{ f_{j}\right\} $ and $\left\{ g_{j}\right\} $
are real functions of time (henceforth referred to as filters) which
must be chosen such that the correlation functions of Eqs. (\ref{eq:kernel_etaeta})-(\ref{eq:zero_correlations})
are satisfied. $F_{j}$ and $G_{j}$ are either 1 or the imaginary unit $i$ and are
also chosen to be consistent with the correlation functions, and the
$\left\{ x_{j}\right\} $ are real valued white Gaussian uncorrelated
noises.

\subsection{Orthogonal Decomposition\label{sec:general}}

The form used above has the benefit that it is possible, if desired,
to decompose each noise into orthogonal components that are correlated
with only one other component \citep{mccaul2018driving,lane2020exactly}.
This orthogonality can, e.g., be achieved by expressing the noises
as

\begin{equation}
\eta(t)=\int_{-\infty}^{\infty}dt^{\prime}f_{1}(t-t^{\prime})x_{1}(t^{\prime})+\int_{-\infty}^{\infty}dt^{\prime}f_{2}\left(t-t^{\prime}\right)\left[x_{2}\left(t^{\prime}\right)+ix_{3}\left(t^{\prime}\right)\right]\label{eq:noise-eta-eta}
\end{equation}
\begin{equation}
\nu(t)=\int_{-\infty}^{\infty}dt^{\prime}g_{1}\left(t-t^{\prime}\right)\left[ix_{1}\left(t^{\prime}\right)+x_{4}\left(t^{\prime}\right)\right]+\int_{-\infty}^{\infty}dt^{\prime}g_{2}(t-t^{\prime})\left[x_{3}(t^{\prime})+ix_{2}(t^{\prime})\right],\label{eq:noise-nu-eta}
\end{equation}
While it is possible to add an arbitrary number of terms of the appropriate
form containing pairs of noises as is done here, we consider no more
than one such term in the expansion of $\eta(t)$ and up to two in
$\nu(t)$, since this restricts the number of necessary white noises
to the minimum possible number. We emphasise that while this does
represent a loss of generality compared to Eqs. (\ref{eq:general_eta}) and (\ref{eq:general_nu}),
there are three benefits. First, autocorrelative and cross-correlative
components of the noise can be immediately identified by their structure,
with, e.g., the first term of Eq. (\ref{eq:noise-eta-eta}) being
autocorrelative while the second term is cross-correlative. Second,
the noise can be decomposed into orthogonal components which are co-correlated
with only one other component. For example, the term involving $f_{2}$
is correlated only with the term involving $g_{2}$ and no other terms.
And third, forming complex noise from pairs of real noises ensures
that their autocorrelation vanishes by construction. This is especially
useful for the $\nu$ noise which has zero self-correlation.

The choice of filters $f_{1}$, $f_{2}$, $g_{1}$ and $g_{2}$ is
then made by relating the expectation values of the noises to the
appropriate correlation functions, Eqs. (\ref{eq:kernel_etaeta})-(\ref{eq:zero_correlations}),
and taking Fourier transforms (indicated by a tilde). In particular,
\begin{equation}
\tilde{K}_{\eta\eta}(\omega)=\tilde{f}_{1}(\omega)\tilde{f}_{1}(-\omega).\label{eq:fourier_K_etaeta}
\end{equation}
Note that $\tilde{f}^{*}(\omega)=\tilde{f}(-\omega)$ for any real
function $f(t)$. Since $K_{\eta\eta}(t)$ is real and even, its Fourier
transform is also real and even, so $\tilde{K}_{\eta\eta}(\omega)=\tilde{K}_{\eta\eta}(-\omega)$,
and thus it is convenient to choose $\tilde{f}_{1}(\omega)$ to be
real, hence
\begin{equation}
\tilde{K}_{\eta\eta}(\omega)=\tilde{f}_{1}(\omega)^{2}\quad\Rightarrow\quad\tilde{f}_{1}(\omega)=\sqrt{\tilde{K}_{\eta\eta}(\omega)},\label{eq:filter-eta-eta}
\end{equation}
thus specifying the autocorrelative filter, $\tilde{f}_{1}(\omega)$.

The correlation between $\eta$ and $\nu$, $K_{\eta\nu}(t)$ of Eq.
(\ref{eq:fgR}), requires that the following constraint in Fourier
space be satisfied:
\begin{equation}
\tilde{f}_{1}(\omega)\tilde{g}_{1}(-\omega)+2\tilde{f}_{2}(\omega)\tilde{g}_{2}(-\omega)=\tilde{R}(\omega),\label{eq:fgR}
\end{equation}
where $R(t)=-iK_{\eta\nu}(t)$ [Eq. (\ref{eq:kernel_etanu})]; note
that $R(t)$ is a real function. Derivations of the Fourier transforms
$\tilde{K}_{\eta\eta}(\omega)$ and $\tilde{K}_{\eta\nu}(\omega)$
and their properties are provided in Appendices \ref{appendix:Ketaeta}
and \ref{appendix:Ketanu}. The three filters $\tilde{g}_{1}(\omega)$,
$\tilde{f}_{1}(\omega)$ and $\tilde{f}_{2}(\omega)$ are determined
by only a single condition [Eq. (\ref{eq:fgR})], and hence their
full specification is subject to different possible choices, some
of which we now discuss.

\subsubsection{\textit{Delta} Scheme\label{subsec:Delta-Scheme}}

Choosing $g_{1}$ to be zero and $g_{2}(t)$ to be a $\delta$ function,
gives
\begin{equation}
{f}_{2}(t)=-\frac{i}{2}{K}_{\eta\nu}(t)
\end{equation}
\begin{equation}
{g}_{2}(t)=\delta(t).
\end{equation}
This choice can be reversed by switching the $\delta$ function around.
For obvious reasons, we refer to this as the \textit{delta} choice;
it was made in previous work \citep{mccaul2018driving}.

\subsubsection{\textit{Constrained} choice\label{subsec:Constrained-choice}}

Taking the constraint Eq. (\ref{eq:fgR}) and setting $\tilde{f}_{2}$
and $\tilde{g}_{2}$ to be zero, this becomes a decomposition with
$\tilde{f}_{1}$ given by Eq. (\ref{eq:filter-eta-eta}) and $\tilde{g}_{1}$
given by
\begin{equation}
\widetilde{g}_{1}(\omega)=\frac{\tilde{R}(-\omega)}{\sqrt{\tilde{K}_{\eta\eta}(\omega)}}.\label{eq:filter-eta-nu}
\end{equation}
We refer to this as the \textit{constrained} choice, since the two
filters $\tilde{f}_{1}(\omega)$ and $\tilde{g}_{1}(\omega)$ are
fully constrained (defined) with no flexibility.

\subsubsection{\textit{Like} Scheme\label{subsec:Like-Scheme}}

In a similar fashion, $\tilde{g}_{1}(\omega)$ can be set to zero
instead of $\tilde{f_{2}}(\omega)$ and $\tilde{g}_{2}(\omega)$,
in which case Eq. (\ref{eq:fgR}) becomes
\begin{equation}
\tilde{f}_{2}(\omega)\tilde{g}_{2}(-\omega)=\frac{1}{2}\tilde{R}(\omega).
\end{equation}
A possible choice for $\tilde{f}_{2}(\omega)$ and $\tilde{g}_{2}(\omega)$
is to require that $\tilde{f}_{2}(\omega)=\tilde{g}_{2}(-\omega)$
such that
\begin{equation}
\tilde{f}_{2}(\omega)=\sqrt{\frac{1}{2}\tilde{R}(\omega)}=\sqrt{-\frac{i}{2}\tilde{K}_{\eta\nu}(\omega)},
\end{equation}
with $\tilde{g}_{2}(\omega)$ simply given by sending $\omega\rightarrow-\omega$
on the right hand side. For obvious reasons, we refer to this choice
as the \textit{like} choice: it has been used by us previously \citep{lane2020exactly}.

\subsubsection{\textit{Reduced} Scheme\label{subsec:Reduced-Scheme}}

Any combination of the \textit{like} and \textit{constrained} choices
will also be allowed, since they would satisfy the general definitions
of the noises Eqs. (\ref{eq:noise-eta-eta}) and (\ref{eq:noise-nu-eta}).
We introduce a set of filters $\tilde{f}_{1}$, $\tilde{f}_{2}$,
$\tilde{g}_{1}$ and $\tilde{g}_{2}$ which utilise both of the above
choices via the introduction of an auxiliary mixing function $\tilde{A}(\omega)$,
\begin{equation}
\tilde{f}_{1}(\omega)=\sqrt{\tilde{K}_{\eta\eta}(\omega)}\label{eq:f1}
\end{equation}
\begin{equation}
\tilde{f}_{2}(\omega)=\sqrt{\frac{1}{2}\tilde{A}(\omega)\tilde{R}(\omega)}\label{eq:f2}
\end{equation}
\begin{equation}
\tilde{g}_{1}(\omega)=\frac{\tilde{R}(-\omega)}{\sqrt{\tilde{K}_{\eta\eta}(\omega)}}\left[1-\tilde{A}(-\omega)\right]\label{eq:g1}
\end{equation}
\begin{equation}
\tilde{g}_{2}(\omega)=\sqrt{\frac{1}{2}\tilde{A}(-\omega)\tilde{R}(-\omega)}.\label{eq:g2}
\end{equation}
Here, the mixing function $\tilde{A}(\omega)$ controls which of the
two choices (\textit{like} and/or \textit{constrained}) is being used
at each value of $\omega$, and it is easy to verify that these filters
satisfy Eq. (\ref{eq:fgR}). A similar expression was recently presented
\citep{imai2015fmo}, though due to incorrect definitions of the filters
it was neither general nor correct, as the properties of the Fourier
transforms (see Appendix \ref{appendix:fourier_transforms}) were
not satisfied in any case except for the autocorrelative component
of $\eta$ which is already fully determined. The special cases of
$\tilde{A}(\omega)=0$ and $\tilde{A}(\omega)=1$ correspond to the
\textit{constrained} and \textit{like} choices, respectively.

By examination of the evolution of $\text{Tr}{(\boldsymbol{\rho}(t))}$
[Eq. (\ref{eq:trace_evolution1})], it is clear that the non-Hermitian
(trace non-preserving) dynamics of the stochastic density matrix is
driven solely by $\nu$. The spread of values of the trace will grow
with time, just as the variance of the displacement of a Brownian
walker grows with time, and this spreading requires an ever larger
ensemble of realisations for the average trace to remain close to
unity at late times. We thus try to choose the mixing function $\tilde{A}(\omega)=\{0,1\}$
to reduce the average amplitude of $\nu(t)$, noting that
\begin{equation}
\langle\lvert\nu(t)\rvert^{2}\rangle=\int\frac{d\omega}{2\pi}\left\{ 2\frac{\lvert\tilde{R}\left(-\omega\right)\rvert^{2}}{\tilde{K}_{\eta\eta}(\omega)}\left|1-\tilde{A}(-\omega)\right|^{2}+\lvert\tilde{R}(-\omega)\rvert\lvert\tilde{A}\left(-\omega\right)\rvert\right\} .\label{eq:<nu2>}
\end{equation}
We choose $\tilde{A}(\omega)=0$ when the first term in the integrand
is smaller than the second term; otherwise $\tilde{A}(\omega)$ should
be 1, that is (cf. Ref. \citep{imai2015fmo})
\begin{equation}
\tilde{A}(\omega)=\begin{cases}
\ 0,\qquad\textnormal{when}\quad\lvert\tilde{R}(-\omega)\rvert^{2}/\tilde{K}_{\eta\eta}(\omega)\leq\lvert\tilde{R}(-\omega)\rvert\\
\ 1,\qquad\textnormal{otherwise}.
\end{cases}
\end{equation}
This choice, which can be done individually for every value of $\omega$,
should then significantly reduce the average magnitude of $\nu(t)$,
diminishing the impact of the non-Hermitian dynamics and improving
the convergence of the ensemble average. Thus we refer to this as
the \textit{reduced} choice.

\subsubsection{\textit{Optimised} Scheme\label{subsec:Optimised-Scheme}}

This naturally leads us to choosing the optimal mixing function $\tilde{A}(\omega)$
which truly minimises the average magnitude of $\nu$; this is the
\textit{$\nu-$optimised} choice. Starting with Eq. (\ref{eq:fgR}),
it can be shown that the mixing function must be real and even (see
Appendix \ref{sec:Optimised-Mixing-Function}). By setting the derivative
of Eq. (\ref{eq:<nu2>}) with respect to $\tilde{A}(\omega)$ equal
to zero, we find the $\nu-$optimised mixing function to be
\begin{equation}
\tilde{A}(\omega)=1-\frac{\tilde{K}_{\eta\eta}(\omega)}{4\lvert\tilde{R}(\omega)\rvert}.
\end{equation}
Substituting this $\tilde{A}(\omega)$ into Eqs. (\ref{eq:f2}) and
(\ref{eq:g2}) gives the corresponding filters as
\begin{equation}
\tilde{f}_{2}(\omega)=\sqrt{\frac{\tilde{R}(\omega)}{2}\left(1-\zeta\frac{\tilde{K}_{\eta\eta}\left(\omega\right)}{\lvert\tilde{R}\left(\omega\right)\rvert}\right)}\label{eq:optf2}
\end{equation}
\begin{equation}
\tilde{g}_{1}(\omega)=\zeta\frac{\tilde{R}\left(-\omega\right)}{\lvert\tilde{R}\left(\omega\right)\rvert}\sqrt{\tilde{K}_{\eta\eta}(\omega)}
\end{equation}
\begin{equation}
\tilde{g}_{2}(\omega)=\sqrt{\frac{\tilde{R}(-\omega)}{2}\left(1-\zeta\frac{\tilde{K}_{\eta\eta}\left(\omega\right)}{\lvert\tilde{R}\left(\omega\right)\rvert}\right)}\label{eq:optg2}
\end{equation}
with $\tilde{f}_{1}{\left(\omega\right)}=\sqrt{\tilde{K}_{\eta\eta}(\omega)}$
as before, and $\zeta=1/4$.

An alternative approach would be to minimise $\langle\lvert\eta\left(t\right)\rvert^{2}\rangle+\langle\lvert\nu\left(t\right)\rvert^{2}\rangle$
rather than just the average magnitude of $\nu$, by considering

\[
\langle\lvert\eta\left(t\right)\rvert^{2}\rangle+\langle\lvert\nu\left(t\right)\rvert^{2}\rangle=\int\frac{d\omega}{2\pi}\left\{ \tilde{K}_{\eta\eta}\left(\omega\right)+\left|\tilde{R}(\omega)\right|\left(\left|\tilde{A}(\omega)\right|+\left|\tilde{A}(-\omega)\right|\right)+2\frac{\lvert\tilde{R}\left(\omega\right)\rvert^{2}}{\tilde{K}_{\eta\eta}(\omega)}\left|1-\tilde{A}(-\omega)\right|^{2}\right\} 
\]
for which the minimising mixing function is

\[
\tilde{A}\left(\omega\right)=1-\frac{\tilde{K}_{\eta\eta}\left(\omega\right)}{2\lvert\tilde{R}\left(\omega\right)\rvert},
\]
with its own $\tilde{f_{2}}$, $\tilde{g}_{1}$ and $\tilde{g}_{2}$,
which are defined by the same Eqs. (\ref{eq:optf2})-(\ref{eq:optg2}),
but with $\zeta=1/2$. We refer to this as the $\eta\nu-$\textit{optimised}
scheme.

The derivations of the minimising mixing function for both optimised
choices are presented in Appendix \ref{sec:Optimised-Mixing-Function}.

It is important to stress that minimising the combined magnitude $\langle\lvert\eta\left(t\right)\rvert^{2}\rangle+\langle\lvert\nu\left(t\right)\rvert^{2}\rangle$
will not necessarily minimise the variance of the trace, nor the rate
of its exponential growth. As far as we are aware it is not possible
to analytically minimise the growth of the trace directly, so we are
forced to approach any optimisation via an ansatz, in this case by
introducing the mixing function and making use of the freedom in its
definition. While the optimal mixing functions derived here affect
the properties of the noises as intended, they do not guarantee that
the behaviour of the trace will be modified in the desired way for
all parameters or over all timescales. This approach should be thought
of as an indirect optimisation of the properties of the dynamics.

\subsubsection{Dynamical Rescaling\label{subsec:Dynamical-Rescaling}}

It is possible to go one step further by introducing a dynamical rescaling
of the cross-correlative filters $\tilde{f}_{2}$ and $\tilde{g}_{2}$.,
as was done for the \emph{like} scheme in previous work \citep{lane2020exactly}. This type of scaling was first introduced for autocorrelative coloured noises in Ref. \citep{shao2010rigorous}, and expanded to cross-correlative noises in Ref. \citep{lane2020exactly}.
Since dividing $\tilde{f}_{2}(\omega)$ by an arbitrary $\omega$-dependent
factor $\tilde{\chi}\left(\omega\right)$ and multiplying $\tilde{g}_{2}(\omega)$
by the same factor will leave the correlation $K_{\eta\nu}$ between
$\eta$ and $\nu$ unchanged, we can choose this factor optimally.
However attempting to minimise $\langle\lvert\eta\left(t\right)\rvert^{2}\rangle+\langle\lvert\nu\left(t\right)\rvert^{2}\rangle$
with respect to $\tilde{\chi}(\omega)$ in Fourier space for each
$\omega$ gives the result that $\tilde{\chi}\left(\omega\right)=\pm1,\pm i$,
which is trivial.

As stated above, while this is the $\tilde{\chi}(\omega)$ which minimises
the combined magnitude of the noises, it is more desirable to minimise
the growth of the trace directly. For this reason we consider a similar
scaling in the time domain, instead dividing $f_{2}\left(t\right)$
by a scaling factor and multiplying $g_{2}\left(t\right)$ by that
same number, even though the scaling freedom is most apparent in Fourier
space. We can then choose the scaling factor to minimise the rate
of spreading of $\lvert\textnormal{Tr}\boldsymbol{\rho}\left(t\right)\rvert$.
We do this by sampling the final value of the trace for a range of
scaling factors and minimising the standard error in the mean trace.
Note that where the optimisation of the mixing function $\tilde{A}$
was analytical, choosing this optimal scaling is a numerical procedure.

It is convenient to implement this scaling via the ratio between the
noises generated using $f_{2}$ and $g_{2}$ before any scaling is
applied, denoted here as $\eta_{0}$ and $\nu_{0}$, respectively.
The scaled noises are then obtained from the unscaled noises as $\eta^{new}=\lambda_{\nu\eta}\eta_{0}$
and $\nu^{new}=\nu_{0}/\lambda_{\nu\eta}$, where

\begin{equation}
\lambda_{\nu\eta}=\sqrt{\lambda}\sqrt{\frac{\sum_{n}\lvert\nu_{0}\left(t_{n}\right)\rvert}{\sum_{n}\lvert\eta_{0}\left(t_{n}\right)\rvert}},\label{eq:scale}
\end{equation}
and $\lambda$ is a parameter (to be determined) representing the
desired ratio between $\nu^{new}$ and $\eta^{new}$. Here, the sums
are over a single realisation of the noises in time, adding the value
of the noise at each discrete time, $t_{n}$.

\subsection{\textit{Convex Optimised} Scheme\label{subsec:Convex-Optimised-Scheme}}

It is also possible to optimise the noise generation scheme in a different
manner using the general form of the noises (\ref{eq:general_eta})
and (\ref{eq:general_nu}), without explicitly introducing a mixing
function \citep{stockburger2019variance}. Instead of minimising the
average of the square magnitude of $\nu$ or the sum of square magnitudes
of $\nu$ and $\eta$, the sum of the imaginary parts of $\eta$ and
$\nu$ can be minimised, subject to the correlations, by the method
of convex optimisation. We can reproduce the analytical expression
obtained in Ref. \citep{stockburger2019variance} for the correlations of
the real and imaginary components of the noises $\eta$ and $\nu$
using the following forms,
\begin{equation}
\eta(t)=\int_{-\infty}^{\infty}dt^{\prime}f_{1}\left(t-t^{\prime}\right)x_{1}\left(t^{\prime}\right)+i\int_{-\infty}^{\infty}dt^{\prime}f_{2}\left(t-t^{\prime}\right)x_{2}\left(t^{\prime}\right)\label{eq:stockburger-a}
\end{equation}

\begin{equation}
\nu\left(t\right)=\int_{-\infty}^{\infty}dt^{\prime}g_{1}\left(t-t^{\prime}\right)\left[x_{1}\left(t^{\prime}\right)+ix_{2}\left(t^{\prime}\right)\right].\label{eq:stockburger-b}
\end{equation}
The filters in Fourier space can be written as

\begin{equation}
\tilde{f}_{1}(\omega)=\frac{1-\tilde{C}(\omega)}{\sqrt{1-2\tilde{C}(\omega)}}\sqrt{\tilde{K}_{\eta\eta}(\omega)}\label{eq:stockburger-1}
\end{equation}

\begin{equation}
\tilde{f}_{2}(\omega)=\frac{\tilde{C}(\omega)}{\sqrt{1-2\tilde{C}(\omega)}}\sqrt{\tilde{K}_{\eta\eta}(\omega)}\label{eq:stockburger-2}
\end{equation}
\begin{equation}
\tilde{g}_{1}(\omega)=\sqrt{1-2\tilde{C}(\omega)}\frac{\tilde{R}(-\omega)}{\sqrt{\tilde{K}_{\eta\eta}\left(\omega\right)}},\label{eq:stockburger-4}
\end{equation}
where

\begin{equation}
\tilde{C}(\omega)=\frac{1}{2}\left[1-\left(\frac{4\left|\tilde{R}(\omega)\right|^{2}}{\tilde{K}_{\eta\eta}(\omega)^{2}}+1\right)^{-1/2}\right].\label{eq:stocknurger-last}
\end{equation}

\subsection{Deconvolution for Reduced and Constrained Schemes\label{subsec:Deconvolution}}

Division in Fourier space can introduce troublesome amplification
for frequencies near which the denominator is close to zero \citep{starck2002deconvolution,hansen2002deconvolution}
(see, for example, Eq. (\ref{eq:filter-eta-nu}).) The $\tilde{C}(\omega)$
function in the convex optimised scheme removes explicit divisions
where this would occur and can be implemented as it stands, as it involves only division by $4\left|\tilde{R}(\omega)\right|^{2}+\tilde{K}_{\eta\eta}(\omega){}^{2}$.
The same applies to the $\nu-$optimised and $\eta\nu-$optimised
schemes where the filters remain finite since $\tilde{R}/\lvert\tilde{R}\rvert$
has real and imaginary parts which are bounded by $\pm1$. Thus the
constrained and reduced schemes are the only schemes which include
explicit division by a filter in Fourier space, in this case by $\sqrt{\tilde{K}_{\eta\eta}\left(\omega\right)}$
in Eqs. (\ref{eq:filter-eta-nu}) and (\ref{eq:g1}), so they require
additional care.

This issue of frequency amplification around the zeros of $\sqrt{\tilde{K}_{\eta\eta}\left(\omega\right)}$
can be eased by deconvolution methods. A deconvolution is the inverse
operation to a convolution which can be naively interpreted as division
in Fourier space. In practice, the process is more complex. Even for
two deterministic functions, there is always an issue of division
close to zero, or of rounding errors which can cause numerical instabilities
in the deconvolved signal after taking the inverse Fourier transform
\citep{smith1997scientist,hansen2002deconvolution}. In particular,
these instabilities can depend on properties of the signal such as
its length $t_{max}$ and spacing $\Delta t$, since these affect
the sensitivity of the Fourier transform to small numbers.

We adopt the deconvolution method of Wiener filtering \citep{wiener1949smoothing}
which minimises the mean square error between some desired quantity
$q(t)$ to be determined and its estimate $\hat{q}(t)$. Considering
the signal associated with $q(t)$ to be

\begin{equation}
y(t)=\int dt^{\prime}\ h\left(t-t^{\prime}\right)q\left(t^{\prime}\right)+\xi(t),
\end{equation}
where $h(t)$ is the known response function of $q\left(t\right)$
and $\xi(t)$ is some unknown noise, the estimate of the signal in
the time domain is
\begin{equation}
\hat{q}(t)=\int dt^{\prime}\ w\left(t-t^{\prime}\right)y\left(t^{\prime}\right),
\end{equation}
where we have introduced some ``inverse'' to the response function,
$w(t)$. In Fourier space this becomes

\begin{equation}
\tilde{q}(\omega)=\tilde{W}(\omega)\tilde{Y}(\omega),
\end{equation}
with

\begin{equation}
\tilde{W}(\omega)=\frac{\tilde{H}^{*}(\omega)}{\left|\tilde{H}(\omega)\right|^{2}+\frac{1}{\textnormal{SNR}}}.
\end{equation}
being the Fourier transform of the inverse response function $w(t)$.
Here $\tilde{H}(\omega)$ and $\tilde{Y}(\omega)$ are the Fourier
transforms of $h(t)$ and $y(t)$, respectively. This $\tilde{W}(\omega)$
is known as the Wiener filter and is used as an estimate of $\tilde{H}(\omega)$
with the problematic frequency amplification removed. It arises directly
from minimising the mean square error $\mathbb{E}|\hat{q}(t)-q(t)|^{2}$
\citep{starck2002deconvolution,van2016comparison}. Finally, $\textnormal{SNR}$
is the signal to noise ratio, or, more concretely, it is the ratio
between the mean power spectral densities of the signal and the noise.
Typically for the Wiener filter, the $\textnormal{SNR}$ needs to be estimated
in some way, especially when the form of the noise $\xi(t)$ is not
exactly known \citep{van2016comparison,starck2002deconvolution}, and
is usually chosen to be a constant value such that the signal is guaranteed
to be larger than the noise.

Adopting this method, the division by $\sqrt{\tilde{K}_{\eta\eta}}$
in the constrained and reduced schemes should be replaced with multiplication
by the corresponding Wiener filter,

\begin{equation}
\frac{1}{\sqrt{\tilde{K}_{\eta\eta}(\omega)}}\rightarrow\frac{\sqrt{\tilde{K}_{\eta\eta}(\omega)}}{\tilde{K}_{\eta\eta}(\omega)+\gamma\max_{\omega}\left|\sqrt{\tilde{K}_{\eta\eta}(\omega)}\right|},
\end{equation}
with a signal to noise ratio $\textnormal{SNR}=\left(\gamma\max_{\omega}\left|\sqrt{\tilde{K}_{\eta\eta}(\omega)}\right|\right)^{-1}$
where $\gamma$ is a small parameter. This allows the correction term
to vary depending on the simulation time $t_{max}$, and to stabilise
the division while still remaining small. Note that this is something
of a numerical fix; it will modify the correlation function $\langle\eta\left(t\right)\nu\left(t^{\prime}\right)\rangle$
so that it no longer matches the desired correlation $K_{\eta\nu}\left(t-t^{\prime}\right)$
[Eq. (\ref{eq:kernel_etanu})] exactly, though the introduction of
the small parameter $\gamma$ allows us to control the size of this
deviation.

\subsection{Deconvolution and Causality}

\begin{figure}
\centering{}\includegraphics[width=0.4\linewidth]{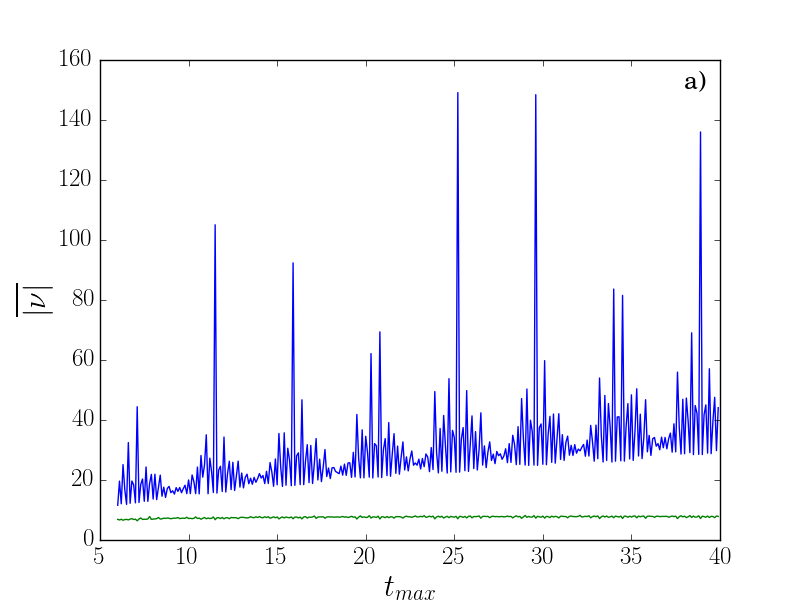}
\includegraphics[width=0.4\linewidth]{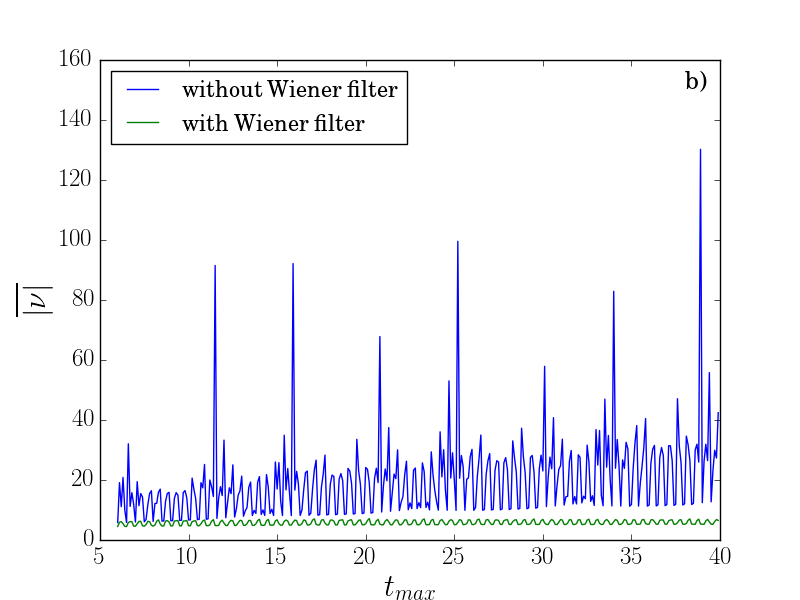}
\caption{Average magnitude of $\nu(t)$ taken across 500 realisations for each
$t_{max}$ for (a) the constrained and (b) reduced schemes with (green)
and without (blue) the Wiener filter using $\gamma=0.01$, $\beta=1$,
$\Delta=1$, $\epsilon=-1$, $\alpha=0.05$ and $\omega_{c}=25$.}
\label{fig:magnitudes_noises_wiener_filter-1}
\end{figure}

\begin{figure}
\centering{}\includegraphics[scale=0.5]{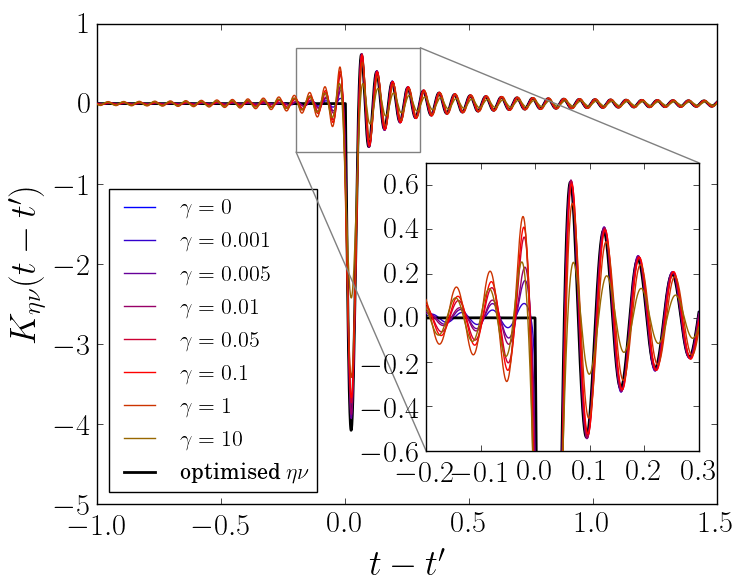}
\caption{The $\eta$-$\nu$ correlation function with different values of the
parameter $\gamma$ in the Wiener filter for the constrained noise
scheme. the $\eta\nu$ optimised scheme overlaps with the desired
correlation $K_{\eta\nu}$ such that $K_{\eta\nu}$ could not be seen,
so it is not shown. $\beta=1$, $t_{max}=12$, $dt=0.01$, $\omega_{c}=25$,
$\alpha=0.05$ for $10^{4}$ realisations. The zoomed inset highlights
the region in which the symmetrisation of the correlation as $\gamma$
increases can be clearly seen.}
\label{fig:etanu_correlation_constrained_wiener-1}
\end{figure}

The instability of the direct Fourier division method can be observed
by investigating the behaviour of the noises for different lengths
of the simulation, $t_{max}$. We compare in Fig. \ref{fig:magnitudes_noises_wiener_filter-1}
the stability of these two schemes with and without the Wiener filter,
by observing the average magnitude of $\nu(t)$ for different values
of $t_{max}$. The application of the Wiener filter to the reduced
and constrained noise schemes improves their stability enormously,
in some cases by as much as an order of magnitude, and significantly
weakens the erroneous dependence of $\nu$ on $t_{max}$, though not
removing it entirely. The trade-off for this improvement is a violation
of the $\eta$-$\nu$ correlation function by introducing a breakdown
of causality, as can be seen in Fig. \ref{fig:etanu_correlation_constrained_wiener-1}.

The application of deconvolution methods thus successfully stabilizes
the $t_{max}$ dependence of $\nu$, decreasing its average magnitude
by reducing the power of frequencies around the singularities in its
spectral density. This improves the convergence and maximum possible
run time of the dynamics, at the cost of weakening causality in the
$\eta$-$\nu$ correlation. Weakening the Heaviside function or removing
it entirely by hand also has this effect of smoothing the $\nu$ noise
and reducing the likelihood of realisations which contain atypically
large values, in turn improving convergence.

While the causality of $K_{\eta\nu}$ is a requirement of the theory,
the introduction of the $\gamma$ parameter gives us a method of deconvolution
for which we can ensure any deviation from the theory is well controlled.

\begin{figure}
\centering{}\includegraphics[scale=0.4]{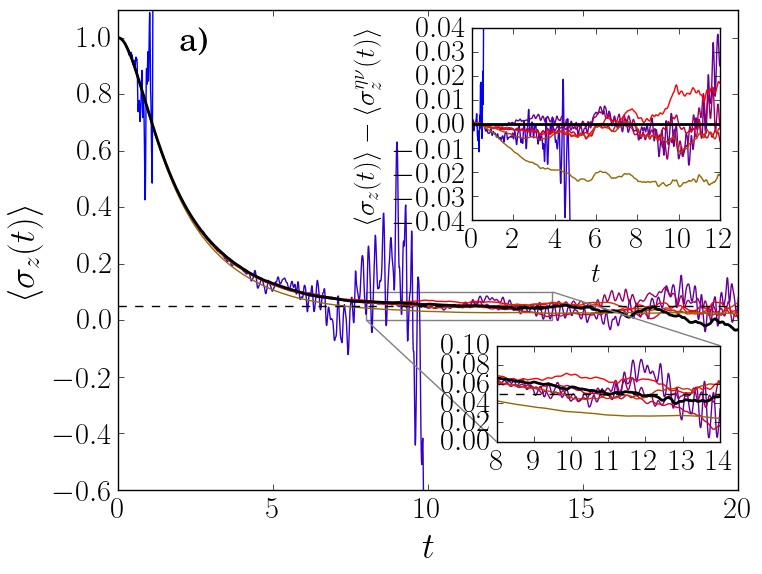}\includegraphics[scale=0.4]{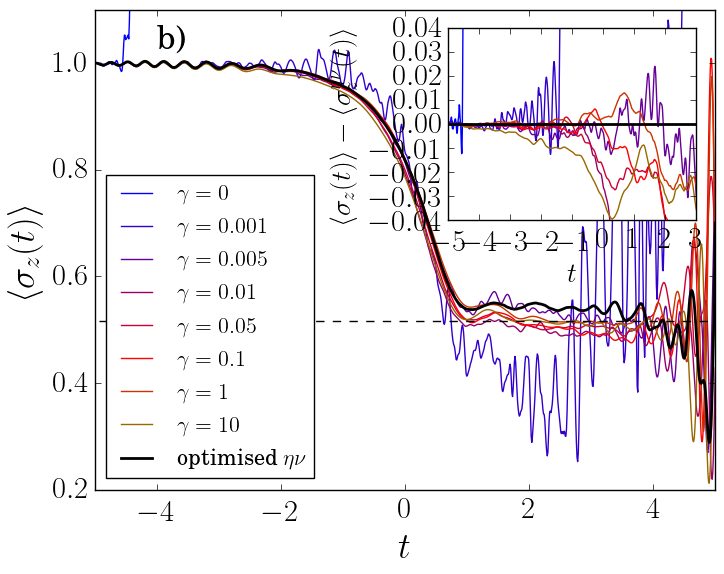}\caption{Comparison of the calculated expectation of the $z$-spin $\langle\sigma_{z}(t)\rangle$
using the constrained scheme with a range of $\gamma$ values to control
deconvolution, and the $\eta\nu-$optimised scheme $\langle\sigma_{z}^{\eta\nu}\left(t\right)\rangle$,
which minimises the combined magnitude of $\eta$ and $\nu$. (a)
The system is initialised with the $z$-spin being 1 and all other
spins being zero and is evolved in time with a constant Hamiltonian,
so that the $z$-spin relaxes to its equilibrium value. The canonical
equilibrium value of 0.05 for the spin, obtained using the imaginary
time evolution methods outlined in \citep{lane2020exactly}, is shown
(dashed line) to confirm the validity of the optimised scheme. The
inset shows the deviation of the $z$-spin for the constrained scheme
from the optimised one. A zoomed-in area at the final stages of equilibration
shows the $z$-spin in detail. (b) A Landau-Zener sweep with a time
dependent Hamiltonian where the system is driven linearly with $\epsilon(t)=5t$,
approaching a known asymptotic limit (dashed line). The system is
initialised at $t=-5$ with the $z$-spin equal to 1 and all other
spins being zero, using the modified Landau-Zener limit of 0.516 to
account for the finiteness of the simulation window as outlined in
\citep{lane2020exactly}. For clarity, data are no longer plotted once
they exceed the vertical scales shown, as the solution becomes unstable
and grows exponentially. $\beta=0.1$, $dt=10^{-2}$, $\Delta=1$,
$\epsilon=-1$, $\alpha=0.05$ and $\omega_{c}=25$ for $10^{6}$
realisations.\label{fig:LZH}}
\end{figure}

We have carried out tests of the above implementation of deconvolution.
In Fig. \ref{fig:LZH} we show the dynamics of the $z$-spin $\langle\sigma_{z}\left(t\right)\rangle$
with a constant Hamiltonian [the relaxation to the equilibrium case,
Fig. \ref{fig:LZH}(a)] and a Landau-Zener sweep [non-equilibrium case, Fig. \ref{fig:LZH}(b)].
The Landau-Zener sweep consists of a linear driving of the form $\epsilon\left(t\right)=\kappa t$,
and has a known analytic solution in the $t\rightarrow\infty$ limit
when the system was initialised in the ground state $\lvert1\rangle$
in the infinite past at zero temperature \citep{zener1932non}. This
limit is \citep{zener1932non,wittig2005landau,rojo2010matrix,saito2007dissipative,orth2013nonperturbative,nalbach2009landau}

\[
\langle\sigma_{z}\rangle_{LZ}=2\exp\left\{ -\frac{\pi\Delta^{2}}{2\hbar\kappa}\right\} -1,
\]
and though originally derived for an isolated spin, it has since been
shown that the same asymptotic behaviour is valid for a dissipative
spin coupled to a harmonic environment at zero temperature, when the
coupling is provided entirely via $\sigma_{z}$ \citep{wubs2006gauging}.
Note that this assumes that the system was initialised in the infinite
past, whereas here it was initialised at $t=-5$. This is taken into
account by modifying the limit appropriately \citep{lane2020exactly},
though there is still some deviation associated with the fact that
the bath is not at zero temperature and that the limit is asymptotic
while the simulation time remains finite.

We expect to recover the canonical equilibrium state (associated with
the constant Hamiltonian) \citep{lane2020exactly} and the Landau-Zener
limit as known solutions at long times in the two cases, and we investigate
the constrained scheme with the Wiener filter for a range of $\gamma$
values, using the $\eta\nu-$optimised scheme which minimises the
sum of magnitudes of $\eta$ and $\nu$ as a reference. Without the
Wiener filter ($\gamma=0$), the constrained scheme diverges almost
immediately for both test cases, whereas for very small $\gamma=0.001$
there is already an improvement, with the accessible simulation time
increasing by $\sim5$ times before $\langle\sigma_{z}\left(t\right)\rangle$
diverges. Note that the behaviour of the $z$-spin after divergence
is omitted for clarity as it oscillates wildly within an exponentially
growing envelope. As $\gamma$ increases to $\sim0.01$ and then to
$\sim0.1$, the constrained schemes begin to converge well, more closely
resembling the $\eta\nu-$optimised scheme result $\langle\sigma_{z}^{\eta\nu}\left(t\right)\rangle$
as can be seen in the insets of Fig. \ref{fig:LZH} where the difference
between them is shown. The statistical convergence is best for larger
values of $\gamma$, most noticeably for $\gamma=10$, though such
a strong Wiener filter introduces a significant deviation from the
$\eta\nu-$optimised scheme and the known solutions, as can clearly
be seen in both the inset and zoomed region in Fig. \ref{fig:LZH}(a).
The same is true in the non-equilibrium Landau-Zener case, Fig.
\ref{fig:LZH}(b), where for smaller $\gamma$ the $z$-spin converges
poorly while for larger $\gamma$ it converges better at the expense
of introducing a deviation from the solution used as a reference.
Thus a compromise value of $\gamma$ must be chosen.

\begin{figure}
\centering{}\includegraphics[scale=0.5]{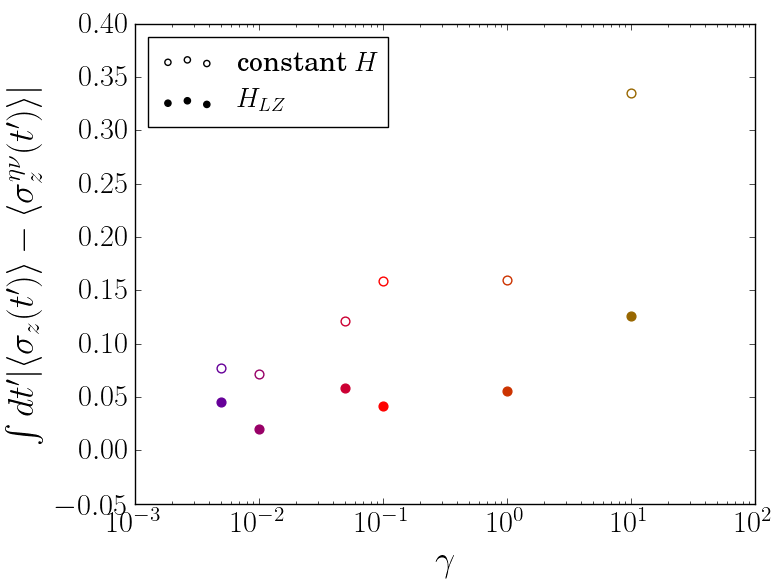}\caption{The total absolute deviation of the dynamics produced using the constrained
scheme for different $\gamma$ from the dynamics produced using the
$\eta\nu-$optimised scheme. Values were calculated using the data
shown in the insets of Figs. \ref{fig:LZH}(a) and (b). Results
for $\gamma=0$ and 0.001 are not shown because the dynamics is diverging
so the deviation is very large. The value of $\gamma$ which minimises
the total deviation for both the constant Hamiltonian (open circles,
Fig. \ref{fig:LZH}(a)) and the Landau-Zener sweep (filled circles,
Fig. \ref{fig:LZH}(b)) is 0.01.\label{fig:measure}}
\end{figure}

The best $\gamma$ value can be chosen by computing the integrated
absolute deviation, $\int dt^{\prime}\lvert\langle\sigma_{z}\left(t\right)\rangle-\langle\sigma_{z}^{\eta\nu}\left(t\right)\rangle\rvert$,
for the data ranges shown in the insets of Fig. \ref{fig:LZH},
presented in Fig. \ref{fig:measure}. This can be thought of as
the total deviation from the $\eta\nu-$optimised scheme within the
region where the convergence of the schemes are comparable, with results
for $\gamma=0$ and 0.001 not shown since they do not remain well
converged on useful timescales. The $\gamma$ which minimises this
quantity is the one with the smallest deviation from the correct dynamics
which we find to be $\gamma=0.01$ for both the constant Hamiltonian
and Landau-Zener cases. By minimising this deviation, we ensure that
the breakdown of causality is well controlled while still managing
to correctly handle the deconvolution and improve the convergence
of the system properties.

\section{Results\label{sec:Results}}

\subsection{Verifying SLN Dynamics with a Quantum Non-Demolition Model}

\begin{figure}
	\includegraphics[width=.45\linewidth]{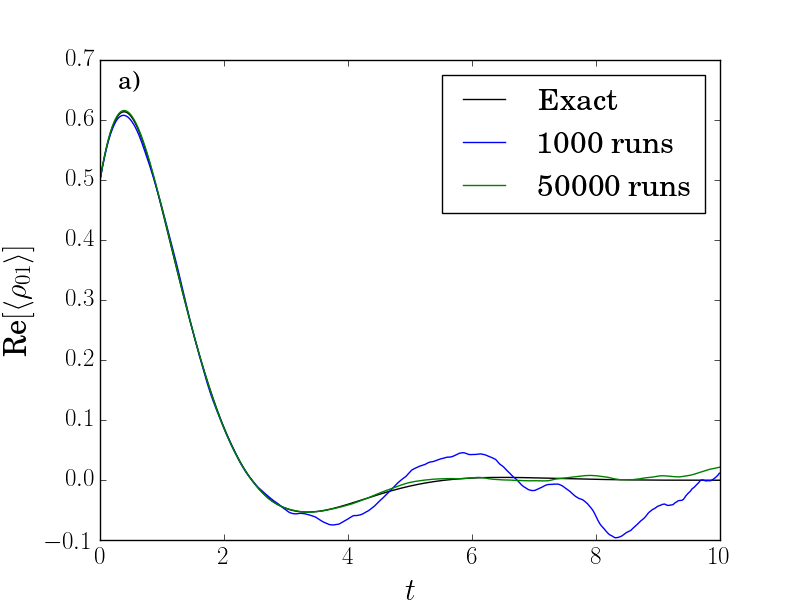}
	\includegraphics[width=.45\linewidth]{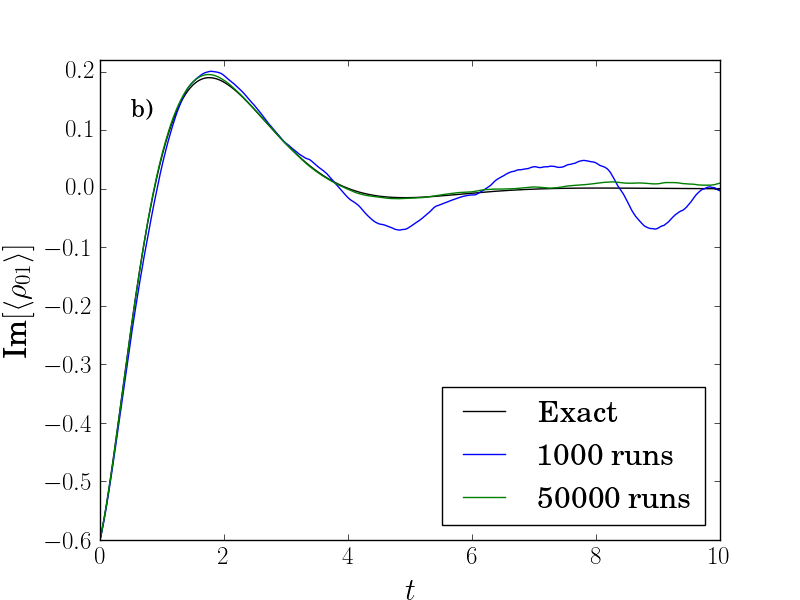}
	\caption{Dynamics of the (01) element of the reduced density matrix according to the quantum non-demolition model, showing the real part $\textnormal{Re}\left[ \langle \rho_{01}(t)\rangle \right]$  in (a), and the imaginary part $\textnormal{Im}\left[\langle  \rho_{01}(t)\rangle \right]$  in (b). The exact solution (black line) is compared to the SLN numerical solutions for 1000 (blue line) and 50000 (green line) realisations, using the $\eta\nu$-optimised scheme with optimal scaling $\lambda=0.5$ (see Sec. \ref{sec:scaling}).}\label{fig:nondemolition_figure}
\end{figure}

In this section, we verify the validity of the SLN equation by comparing the numerical results for $\boldsymbol{\rho}(t)$ simulated using the $\eta\nu-$optimised scheme with the analytical result obtained for a quantum non-demolition model\citep{braginsky1980quantum}. The model considered\citep{yan2016stochastic} is a zero-temperature model with $H_{s} = - \frac{1}{2} \sigma_{z}$, the coupling to the environment is given by $f = \sigma_{z}$, and the environment's correlation function is taken to be $K(t) = \frac{1}{2} \exp \left\{ - 2 |t| + i t \right\}$. Since $f$ and the Hamiltonian commute, the coupling can be thought of as an ideal projective measurement of the open system so as to not disturb its energy\citep{lupacscu2007quantum}. This model can be described exactly by the deterministic master equation\citep{shao2004decoupling,shao1996decoherence}

\begin{equation}
i\frac{d \langle \boldsymbol{\rho}(t) \rangle}{dt} = \left[ H_{s}, \langle \boldsymbol{\rho}(t) \rangle \right] - i C_{r}(t) \left[ {f}, \left[ {f},\langle \boldsymbol{\rho}(t) \rangle \right] \right] + C_{i}(t) \left[ {f}^{2}, \langle \boldsymbol{\rho}(t) \rangle \right],
\label{eq:nondemolition_eq}
\end{equation}
where $C_{r/i} (t) = \int_{0}^{t} d \tau K_{r/i}(t-\tau)$ with  
$K_r(t)=\textnormal{Re}[K(t)]$ and $K_i(t)=\textnormal{Im}[K(t)]$. 
The analytical solution of Eq. (\ref{eq:nondemolition_eq}) is easily found and can be compared to SLN dynamics computed numerically with any of the noise schemes we have considered above, and with  correlations $K_{\eta \eta} = \textnormal{Re}[K(t)]$ and $K_{\eta \nu} = 2i\textnormal{Im}[K(t)]$.
The SLN dynamics using the $\eta\nu-$optimised scheme is shown in Fig. \ref{fig:nondemolition_figure}, along with the analytical solution of Eq. (\ref{eq:nondemolition_eq}), using the initial condition $\langle \boldsymbol{\rho}(t_0) \rangle = 0.5 I + 0.5 \sigma_{x} + 0.6 \sigma_{y}$.

It is clear that the numerical simulation for a stochastic average of 50000 realisations matches the analytical solution for the real and imaginary parts of the density matrix element $\langle \rho_{01}(t) \rangle$ very well. This off-diagonal element is rapidly damped to zero as the environment induces dephasing, with the SLN exhibiting good convergence beyond the initial dephasing and into the equilibrium regime. Additionally, for a much smaller sample of only 1000 realisations, the SLN captures the exact dynamics well for short timescales $t\lesssim2$. Having verified the validity of the SLN equation, in the next section we investigate the numerical efficiency of the noise schemes introduced in Section \ref{sec:noise_generation}.

\subsection{Error Control}

The purpose of the optimisation schemes developed here is to minimise
the typical amplitude of the $\nu$ noise, since it drives the (potentially)
exponential growth of the trace of the stochastic density matrix [Eq.
(\ref{eq:trace_evolution1})]. This should increase the accessible
simulation time (after which convergence is destroyed by numerical
blow up), and reduce the variance of observables.

Without some kind of optimisation, naive choices such as the delta
scheme (Sec. \ref{subsec:Delta-Scheme}) in which one of the
components of $\eta$ or $\nu$ is purely white noise, tend to perform
badly, or even be entirely pathological. The inclusion of white noise
whose variance is one or two orders of magnitude greater than the
trace $\textnormal{Tr}\left(\boldsymbol{\rho}\right)\sim1$ requires
an excessive number of realisations $\gtrsim10^{6}$ for the correlation
functions Eqs. (\ref{eq:kernel_etaeta})-(\ref{eq:zero_correlations})
to converge \citep{mccaul2018driving}, though this by itself does
not guarantee well behaved physical dynamics. Instead, the dynamics
of the trace (or observables) is highly unstable even on very short
timescales, being equally likely to diverge to $+\infty$ as to $-\infty$.
The physical average of such diverging observables will thus tend
to zero as the white noise dominates the dynamics, effectively drowning
out the coupling to the environment via the coloured noise. It is
also clear that any attempt to normalise with the trace when an instability
of this kind has occurred is inappropriate, requiring both division
by zero as well as by very large numbers \citep{stockburger2004simulating,lane2020exactly}.
For these reasons we do not present any data for the delta scheme,
and simply remark that this choice of noise generation scheme is entirely
pathological and should not be used, providing an excellent illustration
that it is not sufficient merely to satisfy the necessary correlation
functions when driving systems using an SLN framework.

\begin{figure}
\centering{}\includegraphics[width=5.8cm]{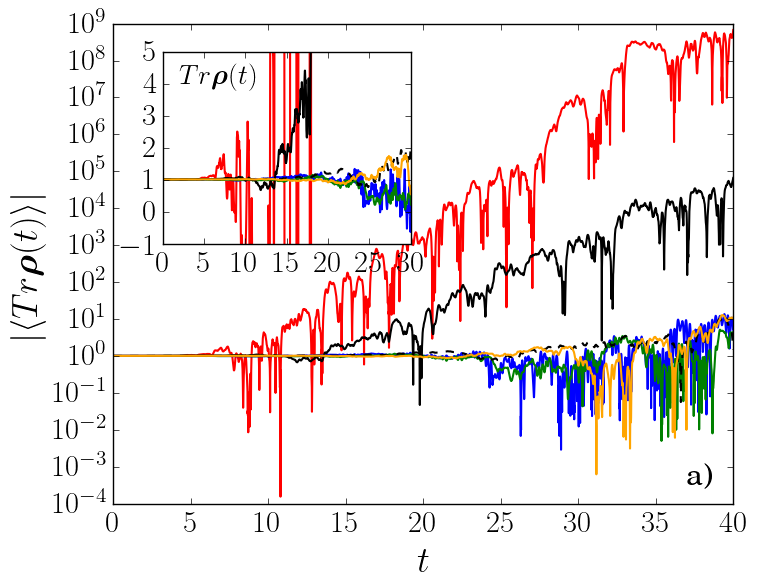}\includegraphics[width=5.8cm]{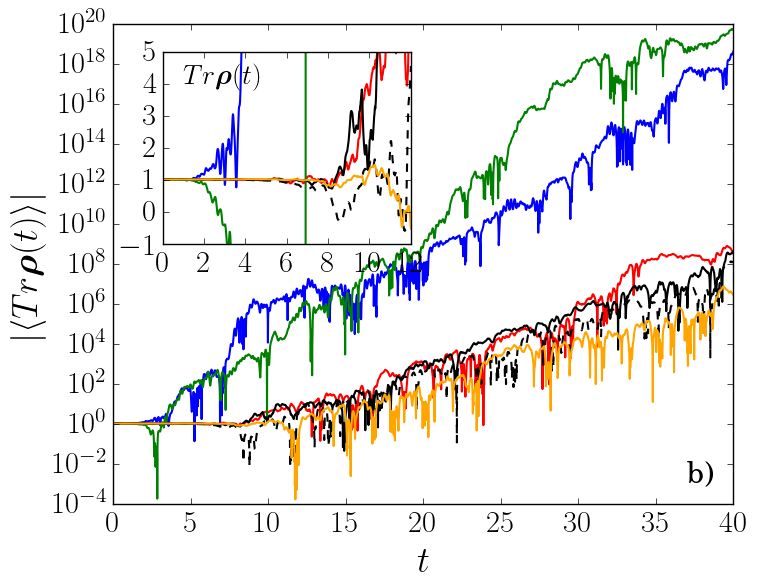}\includegraphics[width=5.8cm]{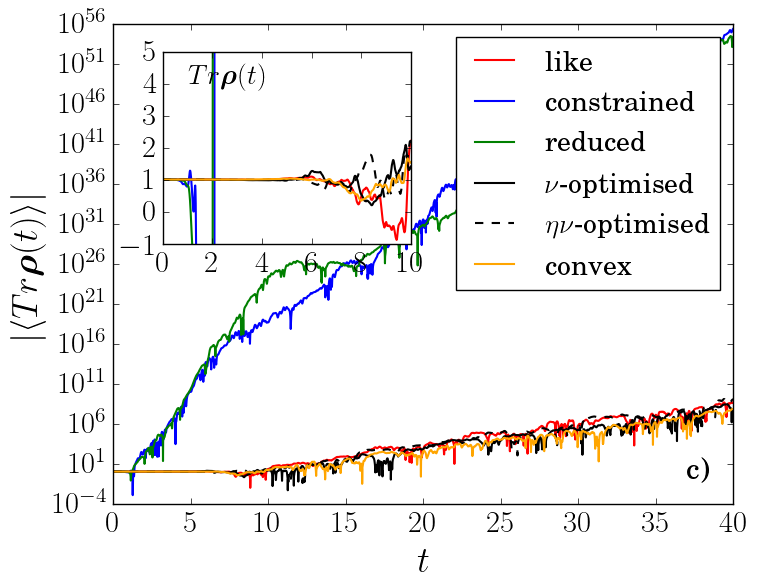}
\centering{}\includegraphics[width=5.8cm]{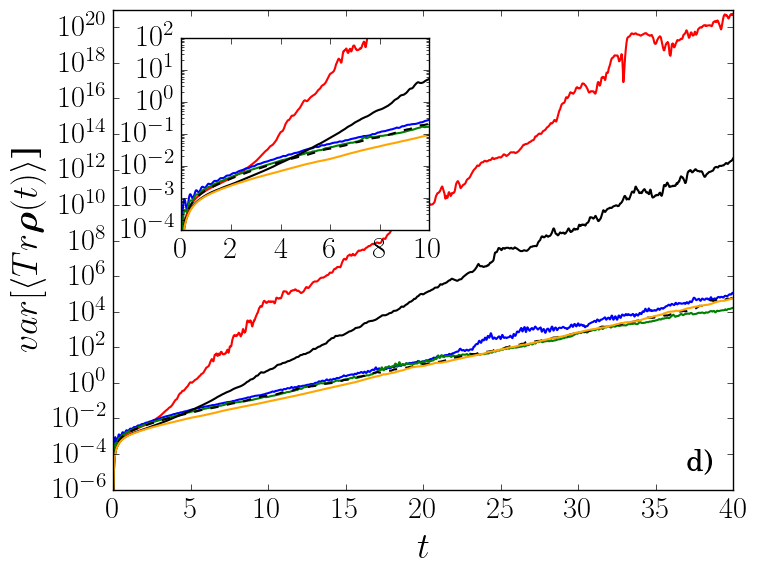}\includegraphics[width=5.8cm]{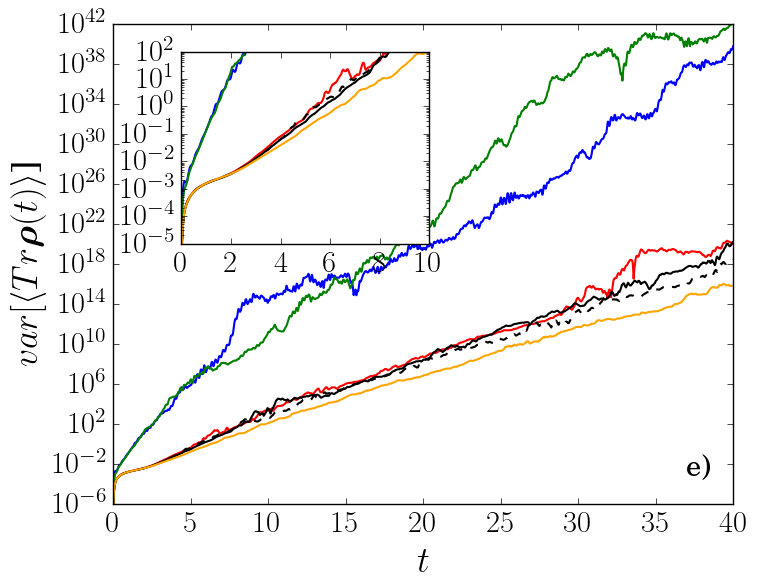}\includegraphics[width=5.8cm]{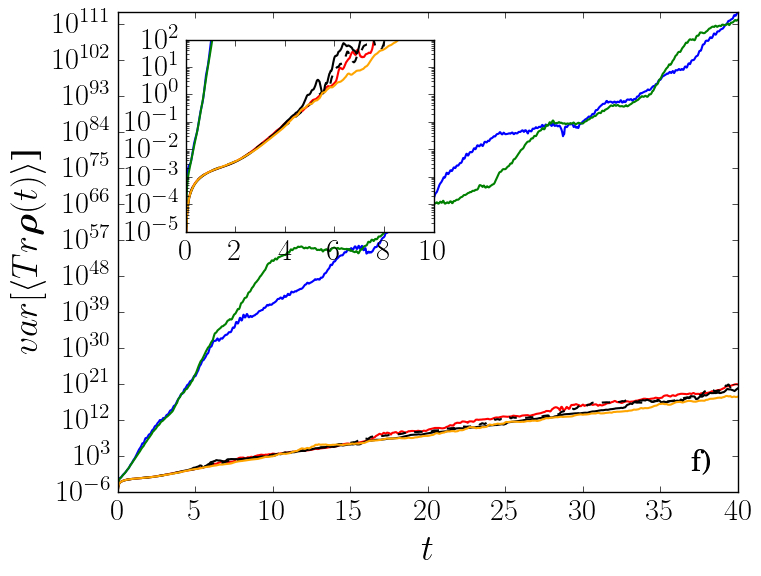}
\centering{}\includegraphics[width=5.8cm]{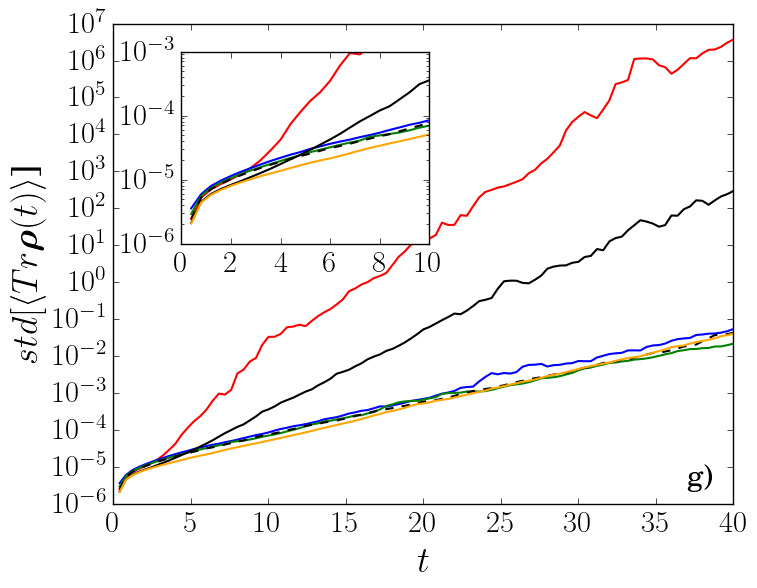}\includegraphics[width=5.8cm]{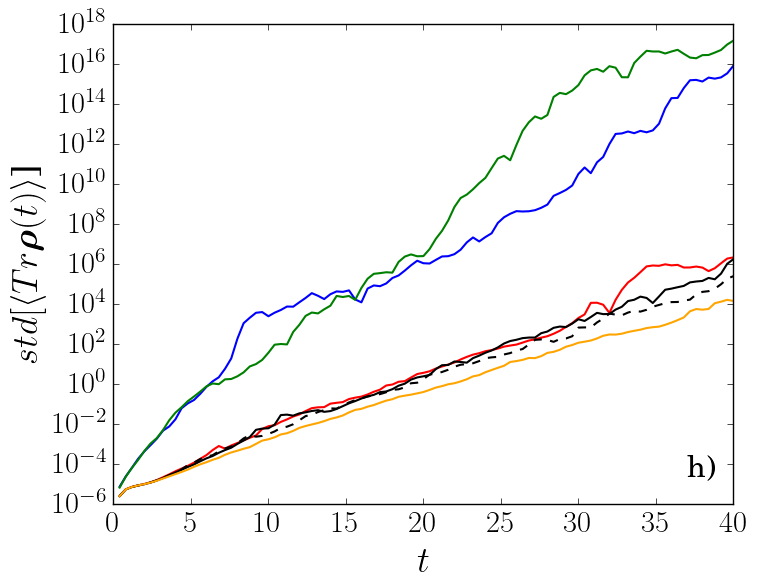}\includegraphics[width=5.8cm]{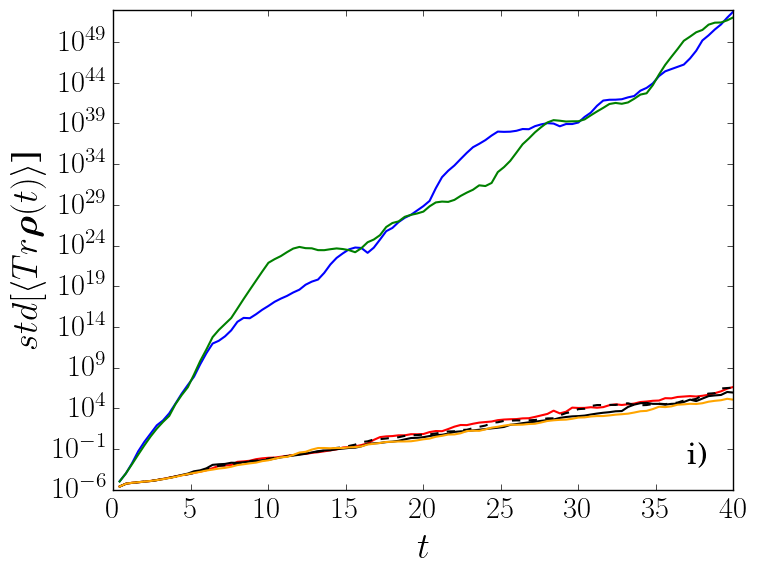}
\caption{(a)-(c): The mean value of the trace, $\langle\textnormal{Tr}\left(\boldsymbol{\rho}\left(t\right)\right)\rangle$,
calculated using different schemes with the system having been initialised
in the state $\lvert1\rangle$ for a constant Hamiltonian. The absolute
value of $\langle\textnormal{Tr}\left(\boldsymbol{\rho}\left(t\right)\right)\rangle$
is shown so that the linear growth on a logarithmic scale is clear.
The insets highlight the timescales on which the simulation is numerically
stable, showing $\langle\textnormal{Tr}\left(\boldsymbol{\rho}\left(t\right)\right)\rangle$
directly; schemes are not shown for timescales beyond which the trace
is clearly diverging. (d)-(f): The variance of the trace and (g)-(i):
the standard error of the mean trace calculated over time windows
which were 100 time steps long. For columns read from left to right,
the inverse temperature increases as $\beta=0.1,1,10$, respectively.
The like (red), constrained (blue) and reduced (green) schemes are
shown, as well as the $\nu-$optimised scheme which minimises $\langle\lvert\nu(t)\rvert^{2}\rangle$
(black solid), $\eta\nu-$optimised scheme which minimises the sum
\textbf{$\langle\lvert\eta(t)\rvert^{2}\rangle+\langle\lvert\nu(t)\rvert^{2}\rangle$}
(black dashed), and the application of the convex optimisation scheme
\citep{stockburger2019variance} as implemented using Eqs. (\ref{eq:stockburger-a})--(\ref{eq:stocknurger-last})
(yellow). All calculations have been done using the same system as
in Fig. \ref{fig:LZH}(a). $\Delta=1$, $\epsilon=-1$, $\alpha=0.05$,
$\Delta t=10^{-2}$, $\omega_{c}=25$ and $10^{5}$ realisations.
No rescaling of the noises was employed.\label{fig:variance}}
\end{figure}

The other schemes all mark a drastic improvement on the naive delta
scheme, as is seen from Fig. \ref{fig:variance}. Recall that the
like scheme (Sec. \ref{subsec:Like-Scheme}) and constrained
scheme (Sec. \ref{subsec:Constrained-choice}) represent the
two distinguishing choices, where $\eta$ and $\nu$ have cross-correlated
orthogonal components, or where all correlations are determined by
$f_{1}$ and $g_{1}$ only, respectively. The optimised choices, barring
convex optimisation, rely on weighting these choices to reduce the
variance of the trace and extend the duration of stable dynamics.

Relative performance of the schemes is illustrated in Fig. \ref{fig:variance},
where we show the mean of the magnitude of the trace $\lvert\textnormal{Tr}\boldsymbol{\rho}\left(t\right)\rvert$
[Figs.  \ref{fig:variance}(a)-(c)], its variance [Figs. \ref{fig:variance}(d)-(f)], and the standard
error of the mean [Figs. \ref{fig:variance}(g)-(i)] for all the schemes at three inverse
temperatures, $\beta=0.1,1,10$. In particular, the performance of the SLN can be quantified via the extent to which the behaviour of the average trace of the reduced density matrix remains constant and close to unity, indicating that the dynamics are physical and well converged, shown in Figs. \ref{fig:variance} (a)-(c). The physical situation is the same
as in Fig. \ref{fig:LZH}(a), where the system is initialised in
the state $\lvert1\rangle$ with the $z$-spin equal to 1 and all
other spins being zero and relaxes towards the equilibrium state associated
with a constant Hamiltonian.

In general, either of the optimised schemes represent a very significant
improvement in the convergence properties and stability of the trace
for the inverse temperatures used, with the growth in the variance
of the trace being drastically reduced [Figs. \ref{fig:variance}(d)-(f)], allowing
an increase in the duration of the stable region [Figs. \ref{fig:variance}(a)-(c)].
However, minimising the typical magnitude of $\nu$ only is found
not sufficient to guarantee this reduction in the variance of the
trace for all temperatures, with the performance of the $\nu-$optimised
scheme only similar to the $\eta\nu-$optimised and convex optimised
schemes at lower temperatures ($\beta=1,10$), but performing much
worse at high temperatures ($\beta=0.1)$.

This is understood by comparing Figs. \ref{fig:variance}(d) and (f) for the variance,
where the $\nu-$optimised scheme and the like scheme both fail for
small $\beta$ while the $\eta\nu-$optimised scheme performs well.
This is caused by the presence of $\textnormal{coth}\left(\frac{1}{2}\beta\hbar\omega\right)$
in $K_{\eta\eta}$ [Eq. \eqref{eq:kernel_etaeta}] which diverges as
$\beta$ becomes small. Since correlation of $\nu$ with $\eta$ enters
via the autocorrelative part of $\eta$ in the reduced scheme, the
amplitude of $\eta$ when generated by the reduced scheme will be
smaller than when generated by the like scheme, as no other noise
component is added to the autocorrelative part. This also explains
why the reduced and constrained schemes perform well for $\beta=0.1$
[Figs. \ref{fig:variance}(a), (d) and (g)]. By accounting for this, the $\eta\nu-$optimised
scheme is an improvement on the $\nu-$optimised scheme despite the
fact that $\nu$ alone is responsible for the intrinsic exponential
growth of the trace; this acts as a reminder that these optimisation
schemes are indirect, in the sense that they do not optimise the properties
of the dynamics of the trace directly.

Accounting for this temperature dependence, the raw $\eta\nu-$optimised
scheme (without any rescaling) and the application of convex optimisation
are comparable, with the benefit that these schemes are universal
rather than depending strongly on the temperature. It is quite fortunate,
as if this were not the case, an investigation of this kind would
have to be performed for every system when selecting a scheme.

\subsection{$\eta\nu-$Optimised Scheme with Rescaling}\label{sec:scaling}

\begin{figure}
\centering{}\includegraphics[scale=0.4]{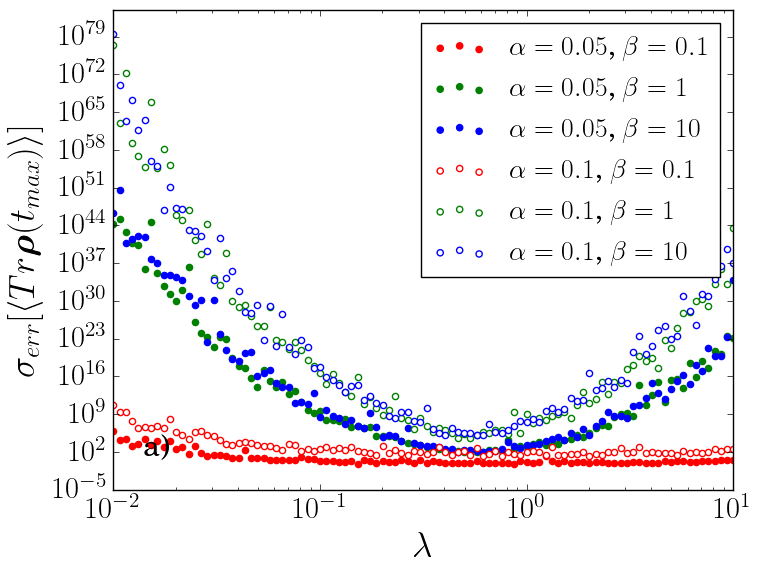}\includegraphics[scale=0.4]{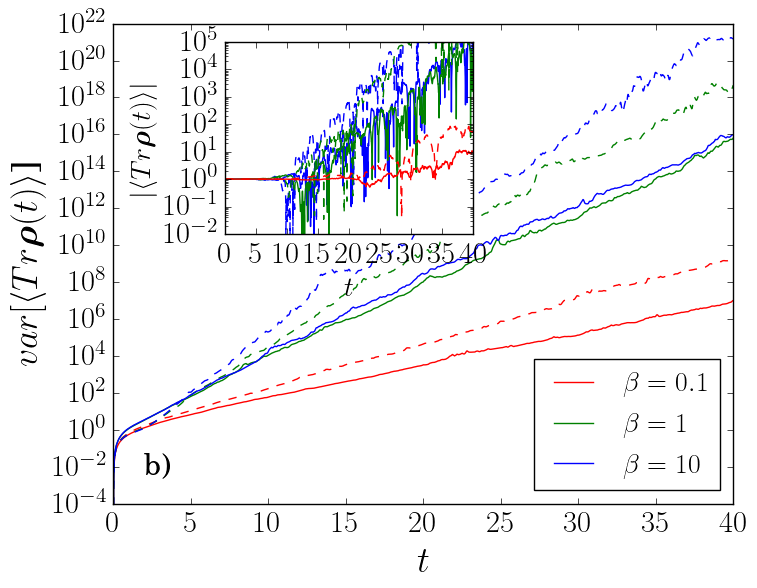}
\caption{(a) The standard error of $\langle\textnormal{Tr}\boldsymbol{\rho}\left(t\right)\rangle$
at its final time step $t_{max}$ as a function of the scaling factor
$\lambda$ for several values of inverse temperature $\beta$ and
coupling strength $\alpha$. For each scaling factor, 1000 runs for
real time dynamics were performed. $t_{max}$ = 40, $dt=10^{-3}$,
$\omega_{c}=25$ and $\Delta=1$, $\epsilon=-1$ for $\alpha=0.05,0.1$
and $\beta=0.1,1,10.$ In this case, the optimum value of $\lambda$
which minimises the growth of the trace is $\approx0.5$. (b) The
variance of the trace having used the $\eta\nu-$optimised scheme
with scaling, with a desired ratio between $f_{2}\left(t\right)$
and $g_{2}\left(t\right)$ of $\lambda=0.5$ (solid lines) for $\beta=0.1,1,10$,
with the mean trace shown in the inset. The convex optimised scheme
(dashed lines) has been reproduced here from Fig. \ref{fig:variance}(d)-(f)
for comparison.\label{fig:scaled}}
\end{figure}

In Fig. \ref{fig:scaled} we apply dynamical scaling to the $f_{2}$
and $g_{2}$ components of $\eta$ and $\nu$ as generated by the
$\eta\nu-$optimised scheme, Eqs. (\ref{eq:optf2})-(\ref{eq:optg2}),
with $\zeta=1/2$. By comparing in Fig. \ref{fig:scaled}(a) the
value of $\lvert\langle\textnormal{Tr}\boldsymbol{\rho}\left(t\right)\rangle\rvert$
at the end of a constant Hamiltonian simulation for a range of rescaling
values $\lambda\in(0.01,10)$ using the procedure of Sec. \ref{subsec:Dynamical-Rescaling},
we find that the optimal value of the scaling is $\lambda=0.5,$ which
we note is the same value obtained previously for the like scheme
\citep{lane2020exactly}. Rescaling the noises with this optimal $\lambda$
using the same parameters as in Fig. \ref{fig:variance}, we find
that the variance of the trace is reduced further, shown in Fig.
\ref{fig:scaled}(b) alongside the convex optimised data from Fig.
\ref{fig:variance} for comparison.

We find that the rescaled $\eta\nu-$optimised scheme is the best scheme for generating noises which minimise the spread and growth (see inset) of the trace for all the schemes considered, at both high and low temperatures.
From a practical perspective, the optimal $\lambda$ can be quickly obtained with only 100 realisations or fewer for each value of $\lambda$, so does not represent a meaningful increase in computational effort.

\section{Discussion and Conclusions\label{sec:Discussion-and-Conclusions}}

In this paper we have developed a number of competing noise generation
schemes, capable of generating complex coloured noises appropriate
for the implementation of the Stochastic Liouville-von Neumann equation.
These noises represent the interaction between the system of interest
and its environment and must satisfy the correlation functions of
Eqs. (\ref{eq:kernel_etaeta}) and (\ref{eq:kernel_etanu}), with the
physical interpretation that averaging over the manifestations of
these noises is equivalent to averaging over all possible behaviours
of the bath. All of the schemes proposed here do satisfy the desired
correlations, but do not otherwise perform equally; that is, the required
sample size for convergence is not uniform between schemes, and nor
is the quality of the subsequent driven dynamics of the reduced system
density matrix. This leads to the important point that there is significant
flexibility in the definitions of the noises, as they are not uniquely
defined by the correlation functions which they must satisfy.

At all stages in this work, great care has been taken to be as transparent
and explicit in the development as possible, in terms of both the
presentation of analytical solutions and the numerical implementation
of the schemes subsequently developed.

Within the general linear filtering ansatz [Eqs. (\ref{eq:general_eta})
and (\ref{eq:general_nu})] we have identified a sub-class of schemes,
which we refer to as orthogonal decompositions \citep{mccaul2018driving},
where the noises are decomposed into components which are correlated only
with one other component (or with themselves), and have the beneficial
property that zero self-correlation can be fulfilled by construction.
There is no limit to the possible choices of the filters with which
these components might be generated from white noise, though we focus
on two such choices for the cross-correlative components between the
$\eta$ and $\nu$ noises: the delta scheme (Sec. \ref{subsec:Delta-Scheme})
where one of the noise components is chosen to be purely white noise,
and the like scheme (Sec. \ref{subsec:Like-Scheme}) where
the filters are chosen so that one is equal to the other with $\omega\rightarrow-\omega$.
The delta scheme represents the worst of the choices, requiring sample
sizes of at least $\sim10^{6}$ for the correlation functions to converge
while still producing unstable dynamics for which the trace rapidly
diverges to $\pm\infty$. This is a prime demonstration that satisfying
the correlation functions alone is not sufficient to guarantee well-behaved
dynamics, or that unrealistically large samples might be required
before the dynamics converges.

Building on an alternative structure for the noises which cannot be
written as an orthogonal decomposition, we followed the arguments
in REf. \citep{imai2015fmo} to develop a scheme which chooses either the
like or constrained scheme (of Secs. \ref{subsec:Like-Scheme}
and \ref{subsec:Constrained-choice}, respectively) at each $\omega$
to reduce the average magnitude of the $\nu$ noise which controls
the spreading of the trace of the reduced system density matrix. Crucially,
by introducing a mixing function $\tilde{A}\left(\omega\right)$ to
blend the schemes and performing a minimisation in Fourier space to
choose it, we were able to ensure that the mixing function was introduced
to the filters in Eqs. (\ref{eq:f2})-(\ref{eq:g2}) correctly such
that the properties of the Fourier transforms of the correlation functions
were maintained. Further, by exploiting these properties and deriving
the Fourier transforms in full, we were able to identify that the
enforcement of causality in the $\eta$-$\nu$ correlation was responsible
for a logarithmic divergence in its Fourier transform. This in turn
causes an amplification of the noise power for frequencies around
the cutoff frequency of the spectral density of the bath, resulting
in weaker convergence than if causality was not required. Fortunately,
by employing the Wiener filter for deconvolutions in Sec. \ref{subsec:Deconvolution},
we were able to parametrise a weakening of causality in cases where
division by zero (or very small numbers) in Fourier space would cause
the spectral densities of the noises to diverge, ensuring that any
deviation from the theory was well controlled while significantly
reducing the $\nu$ noise power.

Going one step further, we explicitly minimised the average amplitude
of both the $\nu$ noise, and the combined amplitudes of the $\eta$
and $\nu$ noises together, in the $\nu-$optimised and $\eta\nu-$optimised
schemes, respectively. We then exploited an additional freedom in
the relative amplitudes of correlated noise components by increasing
the noise power of one component while reducing the noise power of
the other by the same amount so that the correlation functions are
unchanged. We showed that analytic minimisation of the amplitudes
of the noises yields a trivial rescaling, but that direct numerical
minimisation of the standard error of the trace allows us to obtain
an optimal scaling. We emphasise that this scaling is an entirely
independent freedom to the mixing function, and suggest that there
may be many other freedoms and equivalent noise constructions, leaving
space for future work.

Finally, we measured the performance of the aforementioned schemes
along with an alternative optimised scheme (Sec. \ref{subsec:Convex-Optimised-Scheme})
based on convex optimisation \citep{stockburger2019variance} for a
range of inverse temperatures, paying special attention to the properties
of the reduced system trace as a measure of the deviation from the
physical dynamics, as well as its convergence over a set of realisations.
By measuring the variance and standard error of the mean of the trace,
and inspecting how the time at which numerical breakdown occurs varies
for each scheme, we were able to explain why some schemes performed
better at different temperatures than others in terms of competing
noise amplitudes between $\eta$ and $\nu$, and clearly identified
that the re-scaled $\eta\nu-$optimised scheme performed universally
the best out of all the schemes at all temperatures. Remarkably, this
optimisation reduced the variance of the trace by as much as $\sim10^{95}$
at low temperatures and $\sim10^{15}$ at high temperatures. The SLN equation is then compared with the exact solution of a simple quantum non-demolition model, for which near perfect agreement is obtained with statistical convergence extending beyond initial dynamics and into the equilibrium regime.

While comparison with other methods was not within the scope of this paper, we remark that methods which use approximate forms of the bath response function and do not rely so heavily on noises, eg, the hierarchical equations of motions \citep{yan2004hierarchical,zhou2008solving}, achieve well converged results for strong coupling. However, for weaker coupling or arbitrary spectral densities and bath response functions the SLN remains exact, opening an avenue of research for non-Markovian reservoir engineering \citep{breuer_colloqium,TANG}. We hope that this study will stimulate further work in improving the
optimisation of the simulation schemes and consequently will open
avenues for practical numerical simulations of open quantum systems
using SLN and ESLN approaches.

\section*{Acknowledgements}

The first two authors contributed equally to this work.
D.M and M.A.L are supported by the EPSRC Centre for Doctoral
Training in Cross-Disciplinary Approaches to Non-Equilibrium Systems
(CANES, Grant No. EP/L015854/1). Calculations in this paper were performed
using the King\textquoteright s College HPC cluster Gravity.

\appendix

\section*{Appendices}

\numberwithin{equation}{section}

\section{Fourier Transforms}

\label{appendix:fourier_transforms} In deriving the different noise
generation schemes (Sec. \ref{sec:noise_generation}), it was
necessary to use the properties of the Fourier transform of the $\eta-\eta$
correlation function $K_{\eta\eta}(t)$ and the $\eta-\nu$ correlation
function $K_{\eta\nu}(t)$ which we reproduce here.

\subsection{$\tilde{K}_{\eta\eta}(\omega)$}

\label{appendix:Ketaeta} Recalling the definition of $K_{\eta\eta}(t)$
(Eq. (\ref{eq:kernel_etaeta})), its Fourier transform $\tilde{K}_{\eta\eta}(\omega)$
is
\begin{equation}
\tilde{K}_{\eta\eta}(\omega)=\frac{\hbar}{2}\int_{-\infty}^{\infty}dt\int_{0}^{\infty}\frac{d\omega^{\prime}}{\pi}\Lambda(\omega^{\prime})\left[e^{-i(\omega-\omega^{\prime})t}+e^{-i(\omega+\omega^{\prime})t}\right],
\end{equation}
where we have used the shorthand $\Lambda(\omega)=J(\omega)\textnormal{coth}\left(\frac{1}{2}\beta\hbar\omega\right)$
and replaced the cosine with complex exponentials. Using the definition
of the $\delta$ function to remove the time integral,
\begin{equation}
\delta(\omega)=\frac{1}{2\pi}\int_{-\infty}^{\infty}dte^{\pm i\omega t},\label{eq:delta}
\end{equation}
we arrive at the final result,
\begin{equation}
\tilde{K}_{\eta\eta}(\omega)=\hbar\int_{0}^{\infty}d\omega^{\prime}\Lambda(\omega^{\prime})\left[\delta(\omega-\omega^{\prime})+\delta(\omega+\omega^{\prime})\right]=\hbar\Lambda(\lvert\omega\rvert),
\end{equation}
and we can see that $\tilde{K}_{\eta\eta}(\omega)$ is both real,
even and everywhere positive.

\subsection{$\tilde{K}_{\eta\nu}(\omega)$}

\label{appendix:Ketanu} Recalling the definition of $K_{\eta\nu}(t)$,
Eq. (\ref{eq:kernel_etanu}), its Fourier transform $\tilde{K}_{\eta\nu}(\omega)$
is
\begin{equation}
\tilde{K}_{\eta\nu}(\omega)=\frac{1}{2i\pi}\int_{0}^{\infty}\frac{d\omega^{\prime}}{\pi}J(\omega^{\prime})\lim_{\epsilon\rightarrow0^{+}}\int_{-\infty}^{\infty}\frac{d\Omega}{\Omega+i\epsilon}\int_{-\infty}^{\infty}dt\left[e^{-i(\omega+\Omega-\omega^{\prime})t}-e^{-i(\omega+\Omega+\omega^{\prime})t}\right],
\end{equation}
where we have replaced the Heaviside step function with
\begin{equation}
\Theta(t)=\lim_{\epsilon\rightarrow0^{+}}-\frac{1}{2i\pi}\int_{-\infty}^{\infty}d\Omega\frac{e^{-i\Omega t}}{\Omega+i\epsilon},\label{eq:heaviside}
\end{equation}
and replaced the sine with complex exponentials. Again, recognising
the definition of the $\delta$ function, Eq. (\ref{eq:delta}), to remove
the time integral and then using the $\delta$ functions to evaluate the
$\Omega$ integral, we arrive at the following,
\begin{equation}
\tilde{K}_{\eta\nu}(\omega)=-i\lim_{\epsilon\rightarrow0^{+}}\int_{0}^{\infty}\frac{d\omega^{\prime}}{\pi}J(\omega^{\prime})\left[\frac{1}{\omega^{\prime}-\omega+i\epsilon}+\frac{1}{\omega^{\prime}+\omega-i\epsilon}\right].
\end{equation}
We then take the $\epsilon\rightarrow0^{+}$ limit to remove the pole,
\begin{equation}
\lim_{\epsilon\rightarrow0^{+}}\frac{1}{\omega^{\prime}-\omega\pm i\epsilon}=\mathcal{P}\frac{1}{\omega^{\prime}-\omega}\mp i\pi\delta(\omega^{\prime}-\omega),
\end{equation}
($\mathcal{P}$ stands for Cauchy principal value) to obtain
\begin{equation}
\tilde{K}_{\eta\nu}(\omega)=\int_{0}^{\infty}d\omega^{\prime}J(\omega^{\prime})\left[\delta(\omega^{\prime}+\omega)-\delta(\omega^{\prime}-\omega)\right]-\frac{i}{\pi}\fint_{0}^{\infty}d\omega^{\prime}J(\omega^{\prime})\left(\frac{1}{\omega^{\prime}-\omega}+\frac{1}{\omega^{\prime}+\omega}\right)
\end{equation}
\begin{equation}
=-\textnormal{sgn}(\omega)J(\lvert\omega\rvert)-\frac{2i}{\pi}\fint_{0}^{\infty}d\omega^{\prime}\frac{\omega^{\prime}J(\omega^{\prime})}{{\omega^{\prime}}^{2}-\omega^{2}},\label{eq:Ketanu_omega}
\end{equation}
where $\fintop$ also corresponds to Cauchy principal value. Note
that $\tilde{R}\left(\omega\right)=-i\tilde{K}_{\eta\nu}\left(\omega\right)$,
so we immediately see that the real part of $\tilde{K}_{\eta\nu}$
(and the imaginary part of $\tilde{R}$) is odd.

\subsection{The singularity in $\tilde{K}_{\eta\nu}(\omega)$}

\label{appendix:pole} From Eq. (\ref{eq:Ketanu_omega}), we see that
the imaginary part of $\tilde{K}_{\eta\nu}(\omega)$ has an instability
at $\omega^{\prime}=\omega$ that is integrable due to the Cauchy
principle value. By writing $\textnormal{Im}\left[\tilde{K}_{\eta\nu}(\omega)\right]$
as
\begin{equation}
\textnormal{Im}\left[\tilde{K}_{\eta\nu}(\omega)\right]=-\frac{2}{\pi}\fint_{0}^{\omega_{c}}d\omega^{\prime}\frac{{\omega^{\prime}}^{2}f(\omega^{\prime})}{{\omega^{\prime}}^{2}-\omega^{2}},
\end{equation}
where $f(\omega)=\left[1+\left(\frac{\omega}{\omega_{c}}\right)^{2}\right]^{-2}$
and we have used the fact that $J(\omega)$ is zero outside of the range of $\omega$ values $0\leq\omega\leq\omega_{c}$,
we integrate it as follows:
\begin{equation}
\int_{0}^{\omega_{c}}dx\frac{x^{2}f(x)}{x^{2}-\omega^{2}}=\int_{0}^{\omega_{c}}dx\frac{x^{2}\left[f(x)-f(\omega)\right]}{x^{2}-\omega^{2}}+f(\omega)\fint_{0}^{\omega_{c}}dx\frac{x^{2}}{x^{2}-\omega^{2}}.\label{eq:f(x)}
\end{equation}
Only the second term contains the instability which can be handled
as
\begin{equation}
\fint_{0}^{\omega_{c}}dx\frac{x^{2}}{x^{2}-\omega^{2}}=\int_{0}^{\omega_{c}}dx+\omega^{2}\fint_{0}^{\omega_{c}}\frac{dx}{x^{2}-\omega^{2}}=\omega_{c}+\frac{\omega}{2}\ln\left\lvert \frac{\omega_{c}-\omega}{\omega_{c}+\omega}\right\rvert \label{eq:divergence}
\end{equation}
by breaking the Cauchy principal value integral into an integral from
0 to $\omega-\epsilon$ and from $\omega+\epsilon$ to $\omega_{c}$
and seeing that the result is independent of the infinitesimal $\epsilon$.
Hence Eq. (\ref{eq:f(x)}) converges in the Cauchy sense, though a
logarithmic divergence at $\omega=\pm\omega_{c}$ has appeared.

Applying this argument to $\textnormal{Im}\left[\tilde{K}_{\eta\nu}(\omega)\right]$
and simplifying, we arrive at
\[
\textnormal{Im}\left[\tilde{K}_{\eta\nu}(\omega)\right]=-\frac{2}{\pi}\left(\omega_{c}+\frac{\omega}{2}\ln\left\lvert \frac{\omega_{c}-\omega}{\omega_{c}+\omega}\right\rvert \right)f(\omega)
\]
\begin{equation}
+\frac{2}{\pi}\frac{\omega_{c}^{3}}{\omega_{c}^{2}+\omega^{2}}\int_{0}^{1}dx\frac{x^{2}}{\left(1+x^{2}\right)^{2}}\left[\omega_{c}^{2}x^{2}+2\omega_{c}^{2}+\omega^{2}\right].\label{eq:ImK}
\end{equation}
The remaining integrals are then evaluated by relation to the arctangent
to give
\[
\textnormal{Im}\left[\tilde{K}_{\eta\nu}(\omega)\right]=-\frac{2}{\pi}\left(\omega_{c}+\frac{\omega}{2}\ln\left\lvert \frac{\omega_{c}-\omega}{\omega_{c}+\omega}\right\rvert \right)\left[1+\left(\frac{\omega}{\omega_{c}}\right)^{2}\right]^{-2}
\]
\begin{equation}
+\frac{1}{4\pi}\frac{\omega_{c}^{3}}{(\omega_{c}^{2}+\omega^{2})}\left[(6-\pi)\omega_{c}^{2}+(\pi-2)\omega^{2}\right].\label{eq:ImKetanu_omega}
\end{equation}
Thus the imaginary part of $\tilde{K}_{\eta\nu}$ is even and the
real part of $\tilde{R}$ is odd.

The emergence of the logarithmic divergence when $\omega=\pm\omega_{c}$
originates with the presence of the Heaviside step function in the
$\eta\nu$ correlation of Eq. (\ref{eq:kernel_etanu}), which by Eq.
(\ref{eq:heaviside}) and the use of the $\delta$ function introduces
the singularity $\sim\frac{1}{{\omega^{\prime}}^{2}-\omega^{2}}$
in Eq. (\ref{eq:Ketanu_omega}). Since the Heaviside function is an
intrinsic part of the $\eta\nu$ correlation, that is, it was rigorously
derived \citep{mccaul2017partition} rather than being included artificially,
its presence is required by the theory such that removing it any way
would not be formally correct.

\section{Optimised Mixing Function $\tilde{A}\left(\omega\right)$\label{sec:Optimised-Mixing-Function}}

\subsection{Symmetry of $\tilde{A}\left(\omega\right)$}

It is possible to determine some general properties of the real and
imaginary parts of the mixing function $\tilde{A}\left(\omega\right)=\tilde{A}_{1}\left(\omega\right)+i\tilde{A}\left(\omega\right)$
simply from the properties of $\tilde{R}$. Recalling Eq. (\ref{eq:fgR})
coming from $\tilde{K}_{\eta\nu}$ and generalising to arbitrarily
many cross correlative components,

\begin{equation}
\tilde{f}_{1}(\omega)\tilde{g}_{1}(-\omega)+2\sum_{j=2}\tilde{f}_{j}(\omega)\tilde{g}_{j}(-\omega)=\tilde{R}(\omega),\label{eq:fgsum}
\end{equation}
we can then make use of the fact that $\left\{ f_{j}(t)\right\} $
and $\left\{ g_{j}(t)\right\} $ are all real functions. Thus their
Fourier transforms must have even real parts and odd imaginary parts,
since $\tilde{f}^{*}\left(\omega\right)=\tilde{f}\left(-\omega\right)$
for any real function $f(t)$. Then, from Eqs. (\ref{eq:Ketanu_omega})
and (\ref{eq:ImKetanu_omega}), we see that $\tilde{R}\left(\omega\right)=\tilde{R}_{1}\left(\omega\right)+i\tilde{R}_{2}\left(\omega\right)$
has even real part $\tilde{R}_{1}$ and odd imaginary part $\tilde{R}_{2}$.
Using the shorthand $\tilde{f_{j}}=\tilde{f_{j}}^{R}+i\tilde{f_{j}}^{I}$
and $\tilde{g_{j}}=\tilde{g_{j}}^{R}+i\tilde{g_{j}}^{I}$ for the
real and imaginary parts of the filters, we thus have

\begin{equation}
\tilde{R}_{1}\left(\omega\right)=\textnormal{\ensuremath{\sqrt{\tilde{K}_{\eta\eta}\left(\omega\right)}}\ensuremath{\tilde{g}_{1}^{R}\left(\omega\right)}}+2\sum_{j=2}\left[\tilde{f}_{j}^{R}\left(\omega\right)\tilde{g}_{j}^{R}\left(\omega\right)+\tilde{f}_{1}^{I}\left(\omega\right)\tilde{g}_{j}^{I}\left(\omega\right)\right],\label{eq:R1}
\end{equation}
for the real part, where we have used the fact that the real parts
of the filters are even and that the imaginary parts are odd, and
that $\tilde{f}_{1}\left(\omega\right)=\sqrt{\tilde{K}_{\eta\eta}\left(\omega\right)}$
is real. Similarly for the imaginary part we have

\begin{equation}
\tilde{R}_{2}\left(\omega\right)=\sqrt{\tilde{K}_{\eta\eta}\left(\omega\right)}\tilde{g}_{1}^{I}\left(\omega\right)+2\sum_{j=2}\left[\tilde{f}_{j}^{R}\left(\omega\right)\tilde{g}_{j}^{I}\left(\omega\right)-\tilde{f}_{j}^{I}\left(\omega\right)\tilde{g}_{j}^{R}\left(\omega\right)\right].\label{eq:R2}
\end{equation}

For the case we are considering where we include only the $j=2$ term,
and using the filters given by Eqs. (\ref{eq:f1})-(\ref{eq:g2}),
we can determine the symmetry properties of the real and imaginary
parts of the mixing function $\tilde{A}\left(\omega\right)=\tilde{A}_{1}\left(\omega\right)+i\tilde{A}_{2}\left(\omega\right)$.
Since $\tilde{g}_{1}\left(-\omega\right)=\tilde{g}_{1}^{*}\left(\omega\right)$,
the general $\tilde{g}_{1}$ filter requires

\begin{equation}
\left[1-\tilde{A}_{1}\left(\omega\right)-i\tilde{A}_{2}\left(\omega\right)\right]\left[\tilde{R}_{1}\left(\omega\right)+i\tilde{R}_{2}\left(\omega\right)\right]=\left[1-\tilde{A}_{1}\left(-\omega\right)+i\tilde{A}_{2}\left(-\omega\right)\right]\left[\tilde{R}_{1}\left(-\omega\right)-i\tilde{R}_{2}\left(-\omega\right)\right],\label{eq:AR}
\end{equation}
which constrains the real and imaginary parts as

\begin{equation}
\left[\tilde{A}_{1}\left(\omega\right)-\tilde{A}_{1}\left(-\omega\right)\right]\tilde{R}_{1}\left(\omega\right)=\left[\tilde{A}_{2}\left(\omega\right)+\tilde{A}_{2}\left(-\omega\right)\right]\tilde{R}_{2}\left(\omega\right)\label{eq:AR1}
\end{equation}

\begin{equation}
-\left[\tilde{A}_{1}\left(\omega\right)-\tilde{A}_{1}\left(-\omega\right)\right]\tilde{R}_{2}\left(\omega\right)=\left[\tilde{A}_{2}\left(\omega\right)+\tilde{A}_{2}\left(-\omega\right)\right]\tilde{R}_{1}\left(\omega\right)\label{eq:AR2}
\end{equation}
respectively, where we have again used the symmetry properties of
$\tilde{R}$. Assuming that $\tilde{A}_{1}\left(\omega\right)-\tilde{A}_{1}\left(-\omega\right)\neq0$,
then $\tilde{A}_{2}\left(\omega\right)+\tilde{A}_{2}\left(-\omega\right)\neq0$
and dividing Eq. (\ref{eq:AR1}) by (\ref{eq:AR2}) would require
that $\tilde{R}_{1}\left(\omega\right)^{2}=-\tilde{R}_{2}\left(\omega\right)^{2}$
which is obviously incorrect since they are both real. Therefore $\tilde{A}_{1}\left(\omega\right)=\tilde{A}_{1}\left(-\omega\right)$
and then $\tilde{A}_{2}\left(\omega\right)=-\tilde{A}_{2}\left(-\omega\right)$
, i.e. the real part of the mixing function must be even and the imaginary
part must be odd.

Note that the same analysis of Eqs. (\ref{eq:f2}) and (\ref{eq:g2})
results in exactly the same conditions for the mixing function.

\subsection{Minimising magnitude of $\nu(t)$}

Starting with $\nu$ as it is written in Eq. (\ref{eq:noise-nu-eta}),
its magnitude is
\begin{equation}
\langle\lvert\nu(t)\rvert^{2}\rangle=\int\frac{d\omega}{2\pi}\tilde{K}_{\nu\nu^{*}}\left(\omega\right)=\int\frac{d\omega}{2\pi}\left(2\lvert\tilde{g}_{1}\left(\omega\right)\rvert^{2}+2\lvert\tilde{g}_{2}\left(\omega\right)\rvert^{2}\right),\label{eq:nu2}
\end{equation}
where $K_{\nu\nu^{*}}\left(t\right)=\langle\nu\left(t\right)\nu^{*}\left(t\right)\rangle$
and we have made use of Parseval's theorem to remove the exponential
factor associated with the inverse Fourier transform. While it may
at first seem strange that there is no time dependence on the right
hand side, there is no reason why $\langle\lvert\nu\left(t\right)\rvert\rangle^{2}$
should not be stationary. In fact, this apparent stationarity is a
direct consequence of the form of the noises Eqs. (\ref{eq:noise-eta-eta}) and (\ref{eq:noise-nu-eta})
containing time differences in the filters. Substituting in the above
expression Eqs. (\ref{eq:g1}) and (\ref{eq:g2}) and making use of
the fact that $\tilde{K}_{\eta\eta}\left(\omega\right)$ is real and
even, that the real part of $\tilde{R}\left(\omega\right)=\tilde{R}_{1}\left(\omega\right)+i\tilde{R}_{2}\left(\omega\right)$
is even while the imaginary part is odd, we obtain Eq. (\ref{eq:<nu2>})
for $\langle\lvert\nu(t)\rvert^{2}\rangle$. We have also used the
fact that the magnitude of a complex function whose real and imaginary
parts are either even or odd is always real, even and positive. The
aim now is to minimise $\tilde{K}_{\nu\nu^{*}}\left(\omega\right)$
with respect to the real and imaginary parts of the mixing function
$\tilde{A}\left(\omega\right)=\tilde{A}_{1}\left(\omega\right)+i\tilde{A}_{2}\left(\omega\right)$
at each $\omega$ value, where we know that the real part of $\tilde{A}$
should be even and the imaginary part odd. Starting with the real
part,

\begin{equation}
\frac{d\tilde{K}_{\nu\nu^{*}}\left(\omega\right)}{d\tilde{A}_{1}\left(\omega\right)}=\frac{\lvert\tilde{R}\left(\omega\right)\rvert}{\tilde{K}_{\eta\eta}\left(\omega\right)}\left[4\vert\tilde{R}\left(\omega\right)\rvert+\tilde{K}_{\eta\eta}\left(\omega\right)\frac{\tilde{A}_{1}\left(\omega\right)}{\lvert\tilde{A}\left(\omega\right)\rvert}\right]=0,\label{eq:dKnunudA1}
\end{equation}
which yields the following constraint on $\tilde{A}_{1}$ and its
magnitude,

\begin{equation}
\tilde{A}_{1}\left(\omega\right)\left[4\lvert\tilde{R}\left(\omega\right)\rvert+\frac{\tilde{K}_{\eta\eta}\left(\omega\right)}{\lvert\tilde{A}\left(\omega\right)\rvert}\right]=4\lvert\tilde{R}\left(\omega\right)\rvert.\label{eq:b1A1}
\end{equation}

Minimising with respect to the imaginary part of the mixing function
then gives

\begin{equation}
\frac{d\tilde{K}_{\nu\nu^{*}}\left(\omega\right)}{d\tilde{A}_{2}\left(-\omega\right)}=\frac{\lvert\tilde{R}\left(\omega\right)\rvert}{\tilde{K}_{\eta\eta}\left(\omega\right)}\tilde{A}_{2}\left(-\omega\right)\left[4\lvert\tilde{R}\left(\omega\right)\rvert+\frac{\tilde{K}_{\eta\eta}\left(\omega\right)}{\lvert\tilde{A}\left(-\omega\right)\rvert}\right]=0,\label{eq:dKnunu}
\end{equation}
so that $\tilde{A}$ must either be real with $\tilde{A}_{2}\left(\omega\right)=0$,
or the terms within the square brackets must equal zero. If the latter
was true, then Eq. (\ref{eq:b1A1}) would require that $\lvert\tilde{R}\left(\omega\right)\rvert=0$
which is certainly not correct (also, both terms inside the square
brackets are positive), so $\tilde{A}$ is indeed real, $\tilde{A}\left(\omega\right)=\tilde{A}_{1}\left(\omega\right).$
Equation (\ref{eq:b1A1}) then gives

\begin{equation}
\tilde{A}\left(\omega\right)=1-\textnormal{sgn}\left(\tilde{A}\left(\omega\right)\right)\frac{\tilde{K}_{\eta\eta}\left(\omega\right)}{4\lvert\tilde{R}\left(\omega\right)\rvert}.\label{eq:Asgn}
\end{equation}
Since $\tilde{A}<0$ would lead to a contradiction ($\tilde{K}_{\eta\eta}/\lvert\tilde{R}\rvert$
is always positive, so the right hand side would then be positive),
we must conclude that $\tilde{A}$ is a positive function, leading
finally to

\begin{equation}
\tilde{A}\left(\omega\right)=1-\frac{\tilde{K}_{\eta\eta}\left(\omega\right)}{4\lvert\tilde{R}\left(\omega\right)\rvert}.\label{eq:Aopt}
\end{equation}

Substituting this $\tilde{A}$ into the filters of Eqs. (\ref{eq:f2})-(\ref{eq:g2})
gives Eqs. (\ref{eq:optf2})-(\ref{eq:optg2}). It is worth noting
that the division by $\lvert\tilde{R}\left(\omega\right)\rvert$ does
not require serious consideration using the deconvolution procedure
as detailed in Sec. (\ref{subsec:Deconvolution}) since it appears as only a combination $\tilde{R}\left(\omega\right)/\lvert\tilde{R}\left(\omega\right)\rvert$.
Since $\lvert\tilde{R}\rvert\geq\tilde{R}$, this ratio will always
converge without any division by small numbers.

\subsection{Minimising magnitudes of $\eta(t)$ and $\nu(t)$}

We now consider the sum

\[
\langle\lvert\eta\left(t\right)\rvert^{2}\rangle+\langle\lvert\nu(t)\rvert^{2}\rangle=\int\frac{d\omega}{2\pi}\left(\tilde{K}_{\eta\eta}\left(\omega\right)+\tilde{K}_{\nu\nu^{*}}\left(\omega\right)\right)
\]

\begin{equation}
=\int\frac{d\omega}{2\pi}\left(\tilde{f}_{1}\left(\omega\right)^{2}+2\lvert\tilde{f}_{2}\left(\omega\right)\rvert^{2}+2\lvert\tilde{g}_{1}\left(\omega\right)\rvert^{2}+2\lvert\tilde{g}_{2}\left(\omega\right)\rvert^{2}\right),\label{eq:eta2nu2}
\end{equation}
and insert Eqs. (\ref{eq:f1})-(\ref{eq:g2}) to obtain
\begin{equation}
\tilde{K}_{\eta\eta}\left(\omega\right)+\tilde{K}_{\nu\nu^{*}}\left(\omega\right)=\tilde{K}_{\eta\eta}\left(\omega\right)+2\left|\tilde{R}(\omega)\right|\left|\tilde{A}(\omega)\right|+2\frac{\lvert\tilde{R}\left(\omega\right)\rvert^{2}}{\tilde{K}_{\eta\eta}(\omega)}\left|1-\tilde{A}\left(\omega\right)\right|^{2}\label{eq:KetaetaKnunu},
\end{equation}
where we have again used the properties of $\tilde{K}_{\eta\eta}$
and $\tilde{R}$, and the fact that $\tilde{A}$ must be an even function.
The essential difference of the obtained expression from Eq. (\ref{eq:<nu2>})
for $\langle\lvert\nu(t)\rvert^{2}\rangle$ is only in the factor
of two in the last term. Hence, repeating the analysis of the previous
Appendix we obtain

\begin{equation}
\tilde{A}\left(\omega\right)=1-\frac{\tilde{K}_{\eta\eta}\left(\omega\right)}{2\lvert\tilde{R}\left(\omega\right)\rvert}.\label{eq:finalA}
\end{equation}

Note that the same result can be obtained without initially using
the fact that $\tilde{A}$ is an even function, in which case it is
more convenient to write $\tilde{A}$ in the form $\tilde{A}\left(\omega\right)=r\left(\omega\right)e^{i\theta\left(\omega\right)}$.

\bibliographystyle{unsrt}

\end{document}